\documentclass[manuscript,screen]{acmart}
\AtBeginDocument{%
  \providecommand\BibTeX{{%
    \normalfont B\kern-0.5em{\scshape i\kern-0.25em b}\kern-0.8em\TeX}}}

\setcopyright{rightsretained} 
\copyrightyear{2023}
\acmYear{2023}
\acmDOI{XXXXXXX.XXXXXXX}

\acmConference[]{}{}{}

\usepackage{pdflscape}

\begin{document}

\title[Responsible and Inclusive Technology Framework]{Responsible and Inclusive Technology Framework: A Formative Framework to Promote Societal Considerations in Information Technology Contexts}


\author{Juana Catalina Becerra Sandoval}
\email{juana.becerra.sandoval@ibm.com}
\affiliation{
  \institution{IBM Research}
  \streetaddress{1101 Kitchawan Rd}
  \city{Yorktown Heights}
  \state{NY}
  \country{United States}
  \postcode{10598}
}

\author{Vagner Figueredo de Santana}
\email{vsantana@ibm.com}
\affiliation{
  \institution{IBM Research}
  \streetaddress{1101 Kitchawan Rd}
  \city{Yorktown Heights}
  \state{NY}
  \country{United States}
  \postcode{10598}
}

\author{Sara Berger}
\email{sara.e.berger@ibm.com}
\affiliation{
  \institution{IBM Research}
  \streetaddress{1101 Kitchawan Rd}
  \city{Yorktown Heights}
  \state{NY}
  \country{United States}
  \postcode{10598}
}

\author{Lauren Thomas Quigley}
\email{lauren.thomas@ibm.com}
\affiliation{
  \institution{IBM Research}
  \streetaddress{500 North Akard}
  \city{Dallas}
  \state{TX}
  \country{United States}
  \postcode{ 75201}
}

\author{Stacy Hobson}
\email{stacypre@us.ibm.com}
\affiliation{
  \institution{IBM Research}
  \streetaddress{1101 Kitchawan Rd}
  \city{Yorktown Heights}
  \state{NY}
  \country{United States}
  \postcode{10598}
}

\renewcommand{\shortauthors}{Becerra Sandoval, Santana, Berger, Quigley, Hobson}

\begin{abstract}
Technology development practices in industry are often primarily focused on business results, which risks creating unbalanced power relations between corporate interests and the needs or concerns of people who are affected by technology implementation and use.
These practices, and their associated cultural norms, may result in uses of technology that have direct, indirect, short-term, and even long-term negative effects on groups of people and/or the environment.
This paper contributes a formative framework -the Responsible and Inclusive Technology Framework- that orients critical reflection around the social contexts of technology creation and use; the power dynamics between self, business, and societal stakeholders; the impacts of technology on various communities across past, present, and future dimensions; and the practical decisions that imbue technological artifacts with cultural values.
We expect that the implementation of the Responsible and Inclusive Technology framework, especially in business-to-business industry settings, will serve as a catalyst for more intentional and socially-grounded practices, thus bridging the responsibility and principles-to-practice gap.
\end{abstract}




\received{20 February 2007}
\received[revised]{12 March 2009}
\received[accepted]{5 June 2009}

\maketitle


\section{Introduction}

Over the past five years, the ACM Conference on Fairness, Accountability and Transparency (FAccT) has served as an important venue for emergent publications that examine the role of computing technologies in advancing or hampering equity and social justice. Reviews of these publications have noted an abundance of contributions that attend to concerns related to data, models, and to a certain extent human-computer interactions, including efforts to develop tools or methods for conducting bias bounties \cite{Globus-Harris2022}, generating explanations \cite{Lundberg2017, Ribeiro2016}, and improving documentation \cite{Arnold2019, Mitchell2019, Diaz2022}. Pointed critiques, however, have observed comparatively less engagement with socially-grounded modes of inquiry, thus calling for interventions that go beyond technical fixes or ``lab-centric'' approaches to fairness, accountability and transparency~\cite{Gansky2022}, as well as for contributions that address the principles-to-practice gap~\cite{Schiff2020} and attend to real-world scenarios, practices, and stakeholders~\cite{Laufer2022}.

In this paper, we present the theoretical, conceptual, and structural foundation of the \textbf{Responsible and Inclusive Technology Framework (R\&I Framework)}, which we developed to help guide and frame considerations of technology-mediated societal impacts and harms within a business-to-business (B2B) company in the Information Technology (IT) industry. Aligned with conceptualizations of computing technologies as always already social \cite{Green2020, Abbate2022}, the R\&I Framework steers critical inquiry towards investigations of not only the technical, but also the cultural, political, historical, epistemic, and economic factors that impinge on technology creation and use. In this way, the R\&I Framework contributes to a re-orientation of discussions around fairness, accountability, and transparency by focusing on three additional aspects of sociotechnical systems that can be critically interrogated and reflected upon: stakeholders, incentives, and the socio-historical contexts of technology creation and use. Accounting for these aspects jointly, the framework provides a lens for questioning the \textit{who} and \textit{how} of stakeholder participation, the power dynamics and tensions between stakeholders, and the nuances that emerge from contextual specificity.

\begin{figure}[h]
  \centering
  \includegraphics[width=0.7\linewidth]{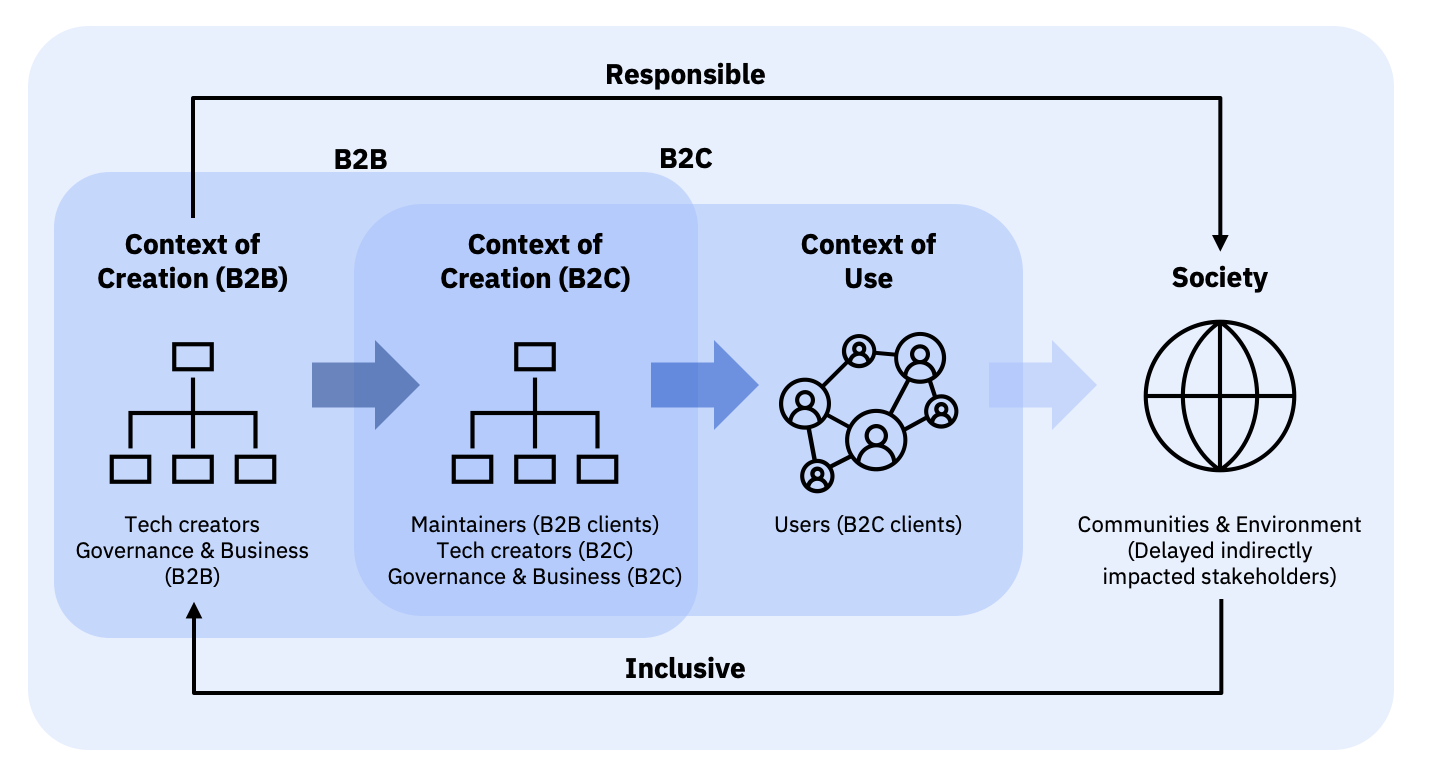}
  \caption{Responsible and Inclusive Framework lenses on accounting for context of creation, context of use, and the responsibility gap potentially present in B2B contexts.}
  \Description{Responsible and Inclusive Framework lenses on accounting for context of creation, context of use, and the responsibility gap potentially present in B2B contexts}
  \label{fig:framework}
\end{figure}

The R\&I Framework is aligned with efforts addressing the dilution of responsibility within large organizations and the responsibility gaps that emerge, in part, due to the long and complex causal chain between technology creation and use~\cite{Goetze2022}. Considering how societal factors are refracted in B2B settings (i.e., settings in which technologists or consultants rarely interact with consumers or end-users and in which the downstream effects of products may not be immediately known), the framework purposefully directs attention to societal stakeholders, as well as indirect and/or delayed social impacts (Figure \ref{fig:framework}). The framework also provides a conceptual framing through which common narratives about technology in social context and assumptions about future uses of technology, especially those that derive from the logics of techno-solutionism \cite{Morozov2013}, techno-universality \cite{Selbst2019}, and techno-beneficence, can be questioned. It nudges towards actions that bridge the gap between the contexts of creation and contexts of use of technology, pushing for participatory engagements and more accountable connectivity across the technology supply chain. 

Although the theoretical and structural framing of the R\&I Framework and the on-the-ground observations that inform it may be of interest to a variety of FAccT audiences, we envision the primary audience for the framework to be people designing, developing, researching, testing, deploying, selling, governing, or otherwise supporting computational technologies and infrastructures within B2B ecosystems.

In the sections that follow, we outline the R\&I Framework and provide details on its theoretical foundation, its structure, its affordances, and its limitations. Section \ref{related_work} presents an overview of related work in FAccT, Science and Technology Studies (STS), participatory and society-centered design, and responsible innovation broadly and provides details on how the framework is informed by and contributes to these bodies of work. Section \ref{research} contextualizes the framework in relation to our larger research initiative, explaining why and how the framework responds to a B2B technology company model in which Artificial Intelligence (AI) is one among many computing technologies. Section \ref{framework} describes the framework in full and introduces stakeholder mapping--specifically the categorization of stakeholders into self, business, and/or society--as a way of surfacing alignments and tensions between the various incentives that impinge on technology. Section \ref{limitations} details limitations and open questions about the framework and its implementation, and, finally, section \ref{conclusion} contains our concluding remarks.

Throughout the paper, and within the framework itself, we follow Stilgoe et al.'s definition of responsible innovation as \textit{a means of taking care of the future through collective stewardship of science and innovation in the present}~\cite{Stilgoe2013}. We are also informed by their characterization of responsible innovation as composed of four dimensions: anticipation, reflexivity, inclusion and responsiveness. However, we situate inclusion as a co-constitutive pillar, rather than a sub-component of responsibility, because inclusion, like responsibility, requires its own modes of anticipatory thinking, reflexivity, and responsiveness. Importantly, we believe inclusion ought to make room for refusal and futurity imagined otherwise, creating opportunities for individuals, community representatives, and activist or civic groups to opt out of participation and reject corporate capture \cite{Hoffman2021}.
\section{Related Work}
\label{related_work}

Throughout the development of the R\&I Framework we relied on theorizations and empirical studies of the `social' in sociotechnical systems and on documented efforts to implement frameworks, methods, and tools to guide responsible and inclusive practices. In what follows, we overview the work that most poignantly informed the framework, including prior work in fairness, accountability, and transparency; science and technology studies; AI and tech ethics; participatory and society-centered design; and responsible innovation and research.

\subsection{Critical Perspectives on Technology and Society}

Historical, sociological, and anthropological analyses of science and technology have contributed to an extensive corpus that questions a priori understandings of technology as neutral, inherently beneficial, and universal. This work includes studies of values, practices, and epistemic norms in science and technology \cite{Daston2007, Shapin2017, Knorr1999, Porter2020, Latour1986}; theorizations of knowledge as situated and informed by social structures \cite{Keller1996, Haraway1990, Harding2003, Cipolla2017}; examinations of the role of science and technology in racial capitalism, empire, and settler colonial projects \cite{TallBear2013, Anderson2006, Hecht2014, AbuElHaj2002, Redfield2000}; and analyses of how science and technology have been leveraged to construct hierarchies of human difference \cite{Hammonds2009, Fullwiley2012, Keel2018, Stern2005}. Interventions looking at computing systems in particular have highlighted how technology can exacerbate inequality, increase conditions of precarity, and augment regimes of coded exposure and surveillance \cite{Benjamin2019, Nakamura2007, Zuboff2019, Noble2018, Atanasoski2019, Magnet2011, Crawford2021, Precarity2020}. 

The question of how to address these technology-mediated social harms has motivated crossdisciplinary and interdisciplinary efforts to bring the above literature to bear in discussions about fairness, accountability, and transparency. Abbate and Dick~\cite{Abbate2022} provide a foundational starting point for this effort. They argue that regardless of whether we are looking at statistical formulations, the technical features of a system, or the consequences of its implementation, we are always in fact looking at \textit{social relations}. These social relations mediate how technology is researched, designed, and developed. For example, in a historical analysis of technical approaches to fairness, Hutchinson and Mitchell~\cite{Hutchinson2019} find that the definitions and approaches that get embedded into technical assessments of fairness are shaped by the civic movements, ethical theories, and value systems of the historical moment that they are responding to. Moore~\cite{Moore2020} similarly finds that decisions about technology are fundamentally rooted in political aspirations and understandings of civic responsibility. Questions such as ``what kind of society does this technology create?'' or ``how can I create the kind of society in which I want to live?'' inform the undertakings of technologists and decision-makers in the IT industry, even if this is not acknowledged explicitly~\cite{Moore2020}. Passi and Barocas~\cite{Passi2019} likewise find that the interactions and acts of negotiation between data scientists and business stakeholders, informed by considerations of profit and capital, also influence how technical problems are formulated, which ultimately affects the kinds of technological futures that get built. 

Social relations also shape how technology is used, a realization that is often missed even within projects that are intended to directly intervene in social systems. Like technologists, ethics owners \cite{Moss2020}, consultants, sales people, and business leaders, the individuals, social groups, and communities who interact with or are affected by technology also have a stake in technology and bring their own set of aspirations, value systems, needs, and concerns. However, studies show that those making decisions about computing systems often have limited knowledge of what is at stake for those affected by sociotechnical systems \cite{voida_2014}. For instance, when Scott et al. \cite{Scott2022} conducted a co-design workshop with jobseekers to explore the implementation of algorithmic systems in Public Employment Services (PES) in Germany, they found that many of the needs and concerns of jobseekers had previously been unaccounted for by PES. Because of this absence of knowledge about the lived realities and the socio-historical contexts in which technologies are used, Katell et al.~\cite{katell2020} argue that investigations of technology-mediated social harms require a situated approach in which community members are involved within systems of accountability. Further, they insist that because ``complex pre-existing social inequities are enabled or reinforced by the implementation of technical systems'' \cite{katell2020} mitigation approaches require an understanding of why these inequalities existed in the first place.

This notion that social relations are central to all aspects of technology, from the moment of ideation all the way to the moment of use or re-use, is foundational to the R\&I Framework. Considering that the social is often not realized, abstracted away, or ignored even in discussions about fairness, transparency, and accountability, the framework's specific focus on stakeholders, incentives, and socio-historical contexts of creation and use (discussed in depth in Section \ref{framework}) aims to bringing those social relations and societal considerations back to the surface, informing the daily work of technologists, designers, researchers, consultants, sellers, and governance decision-makers. In the section that follows, we overview existing frameworks, methods, and tools that similarly attend to the social, clarifying how the R\&I Framework is informed by them, as well as how it contributes and extends their framing.


\subsection{Frameworks, Methods, and Tools for Responsible and Inclusive Technology}

Frameworks that explicitly aim to support  `responsible innovation' vary considerably, depending on whether they center conceptualizations derived from business ethics or rely on broader philosophical understandings of moral responsibility. 
Within frameworks that build on business ethics, responsible innovation is often approached through three main dimensions: (1) avoiding harm, (2) doing good, usually via corporate social responsibility, and (3) implementing governance over the first two dimensions \cite{Voegtlin2017}. Some of these frameworks propose coarse grained principles \cite{Stilgoe2013}, while others focus on specific themes or aspects of innovation such as media \cite{Buhmann2021}, data \cite{Luccioni2022}, licensing \cite{Contractor2022}, workflows \cite{Shneiderman2021}, and trust \cite{Liao2022}. 

Many of these frameworks assume the stance of \textit{personal responsibility}; that is, they are concerned with one's own actions or the actions of the institution one represents. However, arguing for a more capacious engagement with other conceptualizations of moral responsibility, Goetze~\cite{Goetze2022} has proposed an alternative approach: \textit{vicarious responsibility}, which `concerns cases where one agent is responsible for the actions or behaviors of another entity.' This mode of relational responsibility, Goetze explains, occurs in instances where there is a \textit{moral entanglement} between the responsible agent and the entity that performed the actions, which typically happens when one's own agency is somehow implicated in the behavior of the entity. Goetze suggests that vicarious responsibility and moral entanglement are applicable when thinking about technology-mediated harms because of how much decision-making power technology creators have over the affordances and limitations of technology~\cite{Goetze2022}. Another limitation of frameworks for responsible innovation that has been identified pertains to their narrow approach to inclusion. In particular, Ludwig et al.~\cite{Ludwig2020} have highlighted how within business ethics inclusion has typically been thought of as the involvement of the general public in discussions about technology and policy, which sometimes means as little as a request for public comment. 

Aiming to move past these limitations, several research and design frameworks rooted in equity and social justice have unclaimed responsibility as a term and centered their work instead around inclusion \cite{Saja2018}, equity \cite{EXD}, anti-oppression \cite{Smyth2014}, intersectional feminism \cite{Klumbyte2022, Dignazio2020, Collins1986}, and value sensitive design \cite{Hendry2021}. The equityXdesign framework\footnote{https://medium.com/@equityXdesign/racism-and-inequity-are-products-of-design-they-can-be-redesigned-12188363cc6a} developed by Ortiz, Hill, and Molitor \cite{EXD}, for example, offers 3 core beliefs rooted in social justice and critical race theory, 5 human-centered principles, and 4 design processes that help practitioners reflect on and transform their practices. Dimond and Smyth's anti-oppression design framework \cite{Smyth2014} consists of a series of practices and questions that teams can use to prioritize their work, make decisions, and assess their core values. 

In addition to these frameworks, tools for critical reflection have been created and utilized in academic and industry contexts to advance responsible and inclusive practices. Some tools have been created to help reflection around the potential impacts of specific technologies such as biowearables \cite{Minh2021} or mixed reality games \cite{Wetzel2017}. Other tools are technology-agnostic and focus on specific themes like ethical considerations~\cite{Ballard2019}, design-decisions\footnote{https://ddc.dk/tools/toolkit-the-digital-ethics-compass/}, the long-term consequences of technology use \cite{Friedman2012}, and equity and justice\footnote{https://www.artefactgroup.com/case-studies/the-tarot-cards-of-tech/}. Beyond frameworks and tools that emphasize reflexivity, methodological interventions on how to engage with societal stakeholders without imposing extractive dynamics, hierarchies of expertise, and epistemic burden have also been proposed~\cite{Tomasini_Giannini2022-dd, Calderon_Salazar2020-cm, Charlotte_Smith2020-zg, Van_Oorschot2022-ss}. Rooted in participatory action research (PAR) and participatory design (PD)~\cite{Foth2006}, these notably include mutual benefit approaches \cite{Bjorgvinsson2012-kp}, collaborative design \cite{DIgnazio2020-yj}, and speculative design \cite{Harrington2019-pv, Harrington2021, Dunne2013}, to name but a few.  

We created the R\&I Framework inspired by these frameworks, critical reflection tools, and participatory methods, especially those that directly engage with value systems, structural power, and epistemic authority. As a \textit{formative framework}~\cite{Hendry2021} that orients thinking towards societal stakeholders, social impacts, and socio-historical contexts, the R\&I Framework most closely resembles frameworks, methods, and tools that offer and prompt a shared \textit{mode of inquiry}. This is in line with Bietti's argument that modes of inquiry, rather than product-oriented solutions or utilitarian approaches, can help make sense of how technology is intertwined with ``the human, the social, and the political~\cite{Bietti2020}.'' 

The R\&I Framework also extends and contributes to the efforts above by fulfilling the need for a framework that (1) more directly addresses the real-world scenarios, practices, and stakeholders of the B2B context and (2) provides a shared framing and vocabulary that can be used among a variety of people working in B2B contexts, not only technologists but also consultants, sellers, executives, etc. Given this focus, in the R\&I Framework we adopt the concept of `responsible innovation'  because of its familiarity within the business settings we are embedded in and committed to improve; however, we align with Goetze~\cite{Goetze2021} and challenge the traditional notion of personal responsibility by emphasizing the moral entanglements along the technology supply chain. We also directly invoke the concept of inclusion to stress the importance of meaningful participation and socially-grounded considerations of equity and justice--following the steps of the frameworks, methods, and tools above. In the sections that follow, we describe in more detail how we have embedded these considerations into the R\&I Framework.

\section{The Research Initiative}
\label{research}

The R\&I Framework is part of a larger research effort at a multinational large tech company called 
the Responsible and Inclusive Technologies Initiative.
The aim of this initiative is to investigate, develop, transform, and advance enterprise-wide research, development, and business practices to foster a culture that cares about the social impacts of technology and proactively seeks to prevent and address harms. By culture, we refer to shared ideas, norms, and behaviors \cite{Varnum2017}, which in this case includes epistemologies and value-informed decisions. Through mixed methods and embedded research, we create, pilot, and assess frameworks, methods, tools, and other resources that work towards helping researchers, technologists, designers, consultants, sellers, and related practitioners in the company: (1) become more aware of the larger contexts of the technologies and datasets they make, use, support, or sell;
(2) acknowledge and better balance differing incentives and impacts across individual, technical, business, and societal stakeholders; and (3) identify opportunities for proactive action and/or develop mitigation approaches across different levels of organizational capacity and influence. 

The R\&I Framework is one of these resources, and as a result of situated knowledges and practices, it reflects our collective work and personal experiences as researchers with backgrounds in Human-Computer Interaction, STS, History of Science, Bioethics, and Cognitive \& Behavioral Neuroscience. It also reflects our commitments to society-centered design, transformative worldviews, and feminist notions of care. Our familiarity with the enterprise practices and cultures are based on experiences having worked in the company for a few years to decades. 

The framework responds to the company structure and primarily B2B model of operation in which we, and the colleagues we interact with daily, are embedded (See Figure \ref{fig:framework} for a visual representation contrasting B2B and B2C models). B2B contexts present a number of challenges to socially-centered responsible innovation which we have observed in our own experiences and research. For example, most clients in a B2B model are businesses themselves, rather than the end line consumers or users of technology. As a consequence, researchers, developers and other technologists working under this model tend to prioritize stakeholders and conceive of the impacts of their work in relation to business relationships, rather than in relation to the individuals, communities, or environments experiencing the downstream effects of technology. The B2B model also constrains technologists' ability to easily get feedback, measure, and act upon the use of their project or product, in ways that may not apply to smaller or direct-to-consumer/business-to-consumer (DTC/B2C) technology companies. Additionally, pressures related to potential client or product deadlines introduce highly variable timelines that can sometimes shorten time-frames available to engage in critical reflection and forecast the risks of negative impacts for a project. 

Our work is situated within the \textit{specific} context of a large B2B company that offers products and solutions based on a variety of technologies. This means that many of teams we interact with are technologically heterogeneous, with researchers and technologists spanning biology, computer science, engineering, chemistry, physics, design, hardware and software development, and social sciences, to name but a few. Some of our colleagues also hold multiple roles that influence their understanding of responsibility, their awareness of larger company dynamics, and their ability to assert or advocate for change (e.g., practicing scientists who are also managers or directors of larger research portfolios, etc). Moreover, because the company's research endeavors can eventually feed forward into delivered products or services, the larger research enterprise also consists of colleagues in many different roles including business development, sales,  consulting, and legal affairs.  

Given the absence of frameworks or tools oriented at B2B settings observed in Section~\ref{related_work}, as well as the challenges and experiences outlined above, we developed the R\&I framework both in relation to and in reaction to this business model and mode of operation. The B2B context influenced the topics covered, the structure of the framework, and its conceptual approach. For instance, accounting for business values and incentives is central to the framework, as they represent driving forces underlying the kinds of technologies that get made and how research is practiced. Likewise, the framework has a broader focus on technology, rather than solely AI, because many of our research engagements have involved discussions about the complex implications, as well as indirect and/or delayed harms, of other technologies that preceded AI or that are emerging beyond the AI paradigm (e.g., Blockchain, Biotechnology, Quantum Computing). Similarly, the epistemic and organizational breadth of our colleagues directly influenced the scope, depth, specificity, and communication of concepts we had to consider within our framework. We recognize that because of this, not all elements of the framework will be as applicable to people working in a different company or institution, or would be useful to people with different work experiences or expertise. We also refrain from claiming that the R\&I framework is all-encompassing or representative of the best way to approach these issues. Rather, we offer the framework as \textit{a lens} through which the contexts of technology creation and use can be unpacked. 

\section{Responsible and Inclusive Technology Framework}
\label{framework}

The R\&I Framework is structured into two parts: (1) \textbf{the politics of technology creation and use}, in which we propose the categorization of stakeholders into \textit{self}, \textit{business}, and \textit{society} as a way of expanding the notion of who counts as a stakeholder and as a way of more clearly surfacing the conflicts and tensions between the needs and concerns of different stakeholders; and (2) \textbf{the socio-historical contexts of technology creation and use}, in which we outline the logics of \textit{solutionism}, \textit{universality}, and \textit{beneficence} as common assumptions that obfuscate the epistemic, cultural, geographic, and/or political distance between the contexts in which researchers, developers, consultants, and ethics owners work and the lived realities of techno-precarity \cite{Precarity2020} that emanate from how technology is inequitably used and accessed.  

In the sections that follow, we provide more details about each of these components, describing core concerns and considerations, as well as their relevance for the specific audience and uses within a large B2B tech company. We begin with a fictional vignette drawn from a combination of workshops, interviews, and engagements we have had with various technical teams and business units to illustrate the applicability and import of various aspects of the R\&I Framework. Note that the details have been intentionally changed and blended so as to protect the identity of participants and teams, as well as protect the company's and clients' intellectual property and related interests.

\subsection{The Vignette: A Workshop with Team X}

\textit{A fictional team (team X) comprised of researchers, software developers, and business consultants are discussing what it would take to turn their project into a product or service for a potential client Y in the healthcare industry. During a workshop with our team, the group is asked who the stakeholders of the underlying technology and potential use case are. Many people are quick to point to the potential client and their company as key stakeholders given that they are the recipients of the technology and the recipients of the financial gains, respectively.  They also indicate that the client's clients would be stakeholders as they are end-users or consumers of the system; they also identify competitors, who might be developing similar technical capabilities. Future clients are potential stakeholders, especially if the initial project delivery is successful, as it might convince the company to invest further resources to scale up the technology's capabilities. Interestingly, no one lists themselves as stakeholders, despite the fact that (a) the researchers are interested in turning the data into a publication, (b) the development team will be on the hook for translating research code into a working product in a short amount of time, and (c) the consultant may receive commission from securing the deal. Tellingly, no one mentions the people who have to keep up with the demands of the technology after it has been deployed or incorporated into the larger technological infrastructure. In a subsequent workshop, team X begins to identify stakeholder entities outside the direct corporate context (such as ``the government'', ``the media'', ``communities'', ``patients''), but there is little to no specificity or criticality outside an occasional nod towards ``minority groups''. One team member felt that being able to turn their research into a product that hundreds of thousands of people could use was good for society and impactful in and of itself because he assumed it would reduce healthcare costs; another team member expressed discomfort, noticing that they would never be able to see how their machine learning model was used in practice, how well it performed, or if people actually benefited from it.}

\subsection{The Politics of Technology Creation and Use}

Engagements with clients are part of the day-to-day life of technical, governance, and business decision-makers in B2B settings. The nuances of these interactions easily become routine and part of culture. Within the R\&I Framework we call attention to the mundane and highlight how all of these interactions, decisions, and incentive structures are in fact \textit{political}, which we define, following Winner \cite{Winner1980}, as ``the arrangements of power and authority in human associations as well as the activities that take place within those arrangements.'' An objective of the R\&I Framework is to enable technologists, practitioners, consultants, and leaders to reflect on the ways in which the technology they research, create, market, sell, use, or otherwise support embodies, continues, and strengthens relationships of power and authority.  One of the primary ways in which we expose these political influences in technology creation and use is by expanding the notion of who or what could be impacted by a given project or product, calling explicit attention to tensions between stakeholders, incentives, and differences in their prioritization.

\subsubsection{Expanding the Notion of Stakeholders}

The expansion of stakeholders is particularly important in a business-to-business context where consumers or end-users of products aren't easily interacted with and downstream impacts may not be immediately known. In these settings, the notion of stakeholders often stops at people with considerable decision-making power or at people who stand to immediately or directly benefit, materially, financially, and/or reputationally. Although Freeman and Gilbert~\cite{Freeman1992} critiqued this viewpoint in their stakeholder theory work in the 80s and 90s (for example, troubling the idea of shareholders as the priority stakeholders), there is still a tendency to view business stakeholders solely as those with ties to specific economic incentives. There is less discussion of people who may experience indirect gains or outcomes without monetary or corporate-determined value (e.g., network-building, skills, experiences) and even less attention to those who stand to be substantially negatively impacted despite having little if any influence on how products or projects get designed or implemented. When considered, which happens rarely, affected societal stakeholders are typically lumped into monolithic categories, and assigned the role of either receiving education or responding to a survey~\cite{Greene2019}, as opposed to being seen as individuals and communities with a diverse set of experiences and backgrounds, who not only have the knowledge to educate scientists and technologists but also the ability and right to survey and critique technologies based on their lived expertise. 

Our framework addresses narrow and abstract understandings of stakeholders in three ways: (1) it treats the notion of `business' (including technologists and practitioners within a business) as \textit{inseparable} from the social; (2) it rejects the notion of a singular ``public,'' thinking instead about individuals, social groups, communities, and environments as stakeholders with their unique values, interests, needs, concerns, epistemologies, etc.; and (3) it redirects attention to societal stakeholders and social stakes which frequently are de-prioritized by business stakeholders and business stakes. With these three features in mind, we forward the definition of \textbf{stakeholders} as \textit{any person, entity, organization, or community that currently has or has previously had a stake in the work now, or will in the future; a stakeholder is anyone or any thing that is/was/will be directly or indirectly impacted by the work, anyone who directly or indirectly influences the work (or has in the past/will in the future), or anyone who cares about the work (or has in the past/might in the future).} This way of framing stakeholders intentionally makes room for non-human considerations (e.g., environment, animals and living ecosystems, digital and material infrastructures), while also calling attention to nonlinear effects, multiple and often conflicting priorities, and/or temporal delays in potential impact. 

Due to the on-going tendency to prioritize business stakes and corporate logics over societal stakes, as well as the frequent practice of not reflecting on personal stakes within a project,  we posit three categories of stakeholders that ought to be identified and reflected upon within the R\&I Framework: 

\begin{itemize}
    \item \textbf{Self} - The self can be seen as the individual researcher, developer, designer, consultant, decision-maker, practitioner, etc. The self can also be seen as encompassing an entire team, project, or initiative. In essence, the self encompasses those working most closely with the research process, technology, datasets, or product. 
    \item \textbf{Business} -  Intentionally broad, business includes but is not limited to larger groups like research or development departments, business units, enterprise-wide leaders, data providers, clients, partners, and even competitors. 
    \item \textbf{Society} - In support of the idea of multiple heterogenous publics, society encompasses end-users, social groups, communities, civil constituents, legislators, environments, and many other entities, locally, regionally, nationally, and globally.
\end{itemize}

We understand these stakeholder categories as \textit{framing}, rather than definitional categories. This means that someone closely involved in a project (thus fitting within the self) could also be directly impacted by the implementation of the technology as a civil constituent (thus also fitting within society). This also means that stakeholders could fit within different categories depending on the contexts. For example, in some scenarios government institutions may be clients of computing technologies (thus fitting within business), while in other scenarios they may act as regulators (thus fitting within society). 

\subsubsection{Accounting for Incentives, Power Dynamics, and Tensions}

Each stakeholder of a project has its own combination of values and beliefs, interests, incentives, norms, and political and economic power that are up-taken, enforced, or disseminated within sociotechnical systems~\cite{Friedman1996}. Within the R\&I Framework we refer to these combinations as \textit{forces}. These forces carry weight by putting pressure against or towards certain outcomes or by removing pressures to sustain or uphold certain logics, similar to the idea of `force fields' proposed by Stamper \cite{Stamper1993} as part of the Organizational Semiotics theory. Because of contextual dependence, we do not presume that there is anything inherent about these forces, e.g., that certain values are necessarily ‘wrong’ or ‘right’ or that there are always clear delineations between interests.  Importantly, we do not see these forces as separate or necessarily distinct but rather inter-enveloped, inter-personal, and in some instances, co-constituting. For example, data science teams in applied corporate settings are often inherently collaborative, with "project managers, product designers, and business analysts as much a part of applied real-world corporate data science as are data scientists" \cite{Passi2018} - these groups might be simultaneously influenced by self forces in terms of personal interests and associated team research or project objectives, as well as business forces related to successful on-time product or service delivery.  

We also don't see each force as being self-homogeneous - we don't expect values, incentives, or dynamics to be perfectly aligned or always congruent. It is problematic to assume that someone's technical background or their employment at a certain company directly reflects their beliefs, identities, or values; as Terzis~\cite{Terzis2020} has pointed out, one of the most fundamental issues in responsible and ethical computing is the lack of acknowledgment that every agent has subjectivity in a given context, be it the CEO, the project manager, the business analyst, the developer, or the micro-worker. Leidner~\cite{Leidner2006} also mentions that individual differences within the particular cultural unit or team may lead to different behavioral outcomes. Critically reflecting on these forces and the intersections of self, business, and society elucidates differences in decision-making power and prioritization, as well as the \textit{tensions} that emerge from these differences.

The R\&I Framework follows Flyvberg et al.'s notion of tensions as ``power relations [that] are fraught with dubious practices, contestable knowledge and potential conflict''~\cite{Flyvbjerg2012}. These tensions are integral to understanding diverse perspectives and can lend to productive frictions between stakeholders, creating opportunities to interrogate power differentials and infrastructural or systemic barriers within the technology life cycle \cite{stark2009}.  Tensions are also critical sites of interrogation and intervention because, as Flyvbjerg \cite{Flyvbjerg2012} and others have noted, they are particularly susceptible to problematization, which means they are also susceptible to cracks, deconstruction, and ultimately \textit{change}.  For this reason, some have argued that focusing on tensions can help articulate and redirect research priorities within AI ethics, responsible or good tech initiatives, and more \cite{Whittlestone2019}. By making these power dynamics and invisible structures more visible, it may aid in ``assessing impact on people and creating space for reflection and repair'' \cite{EXD}.

Returning to the introductory vignette and team X, the motivations between researchers, developers, and consultants differed along lines of personal fulfillment and technological merit (self) versus time and labor specifications, financial incentives, or even abiding by regulatory requirements (business).  We can also imagine how these incentives might differ from those of client Y, who wants a working product (business), and how further divergent these perspectives might be from those whose livelihoods and health may be impacted by longer-term changes instantiated by the technology, be these end-users or other constituents (society). Because (1) business incentives often ``favor the interests of industry stakeholders instead of the common good''~\cite{Bietti2020}, (2) business strategies can supersede technical considerations such that the most profitable solution is selected over the best technical solution~\cite{cowan1985}, and (3) business models can enable the dilution of responsibility~\cite{Benjamins2021}, deliberate efforts to critically reflect on and balance business and societal forces are foundational to responsible and inclusive practices. 

\subsection{The Socio-historical Contexts of Technology Creation and Use} 

In addition to promoting the identification of a broader set of stakeholders and acknowledging the possibility of competing incentives and motivations, the R\&I Framework forwards an understanding of stakeholders as situated within and informed by a particular socio-historical context. In the vignette above, a member of fictional team X identified `patients' as a stakeholder and hypothesized that the project would benefit patients by reducing healthcare costs. But who are those patients? where do they live? how old are they? what kind of healthcare plans do they have access to? and which barriers do they face when trying to receive care? All of these questions matter because they elucidate in more detail the lived realities of those who will interact with or be affected by the project. 

Also important are the differences between the socio-historical contexts in which the project was researched, designed, developed, commercialized, deployed, and used. Suppose team X's engagement with client Y was a massive success and now they are receiving requests from other clients who want the same product. Could the product perform differently for clients located in other cities, countries, or even continents? Could the results provided by the product be interpreted differently depending on cultural references or ways of knowing? Could the consequences of using the product diverge from what was observed when client Y used it? 

\subsubsection{Narratives and Assumptions about Technology in Social Context}

Informed by the literature outlined in Section \ref{related_work} and our experiences interviewing and conducting workshops with various technical and business teams, we identify within the R\&I Framework three common assumptions that hinder this kind of socio-historical contextualization: techno-solutionism, techno-universality, and techno-beneficence. The framework specifically calls for questioning and critically reflecting on these common assumptions because, as a whole, these assumptions over-simplify social phenomena and present technology in an overly optimistic fashion, thus holding us back from thoroughly forecasting and/or understanding the impacts of technology. Within efforts to engage in responsible and inclusive practices, these assumptions are dangerous especially when they reduce the scope of mitigation strategies and foreclose alternative ways of addressing social problems.

Within the framework, we begin by calling attention to \textbf{techno-solutionism}, which we describe as \textit{the logic wherein all social problems are presented as solvable using technology}~\cite{Morozov2013}. In some cases, techno-solutionism goes as far as to enable the formulation of social problems as best solved when done so using computational tools and methods. Although some problems may be effectively and justifiably addressed through technical means, what our conceptualization of techno-solutionism captures is the \textit{systematic assumption} that this will be the case, even when there is evidence pointing to the contrary. Techno-solutionism is related yet epistemologically distinct from what is often called The Law of the Instrument~\cite{Maslow1966}; while The Law of the Instrument states that everything looks like a nail to a person with a hammer, the logic of techno-solutionism characterizes and molds everything into a nail, just so that the hammer can be used. The problem with techno-solutionism is that it risks diverting resources away from other solutions that may be better suited to the problem. Techno-solutionism may be particularly, and at times problematically, encouraged in B2B settings, where doing business in a `solutions-oriented' way is common~\cite{Woodings2006, Shepherd2000, Agafonow}, where coming up with client-centered technical solutions is rewarded, and where additional information which might be interpreted as extraneous or even problem-oriented is discouraged~\cite{Watts}.

We also caution against \textbf{techno-universality}, which we describe as \textit{the assumption that a technical solution that is applicable and works well in a particular context can be generalized to all other contexts}. For decades, the concept of universality has proven particularly fruitful for physics and other fields that rely on universal, constant variables, as well as for fields that seek to establish fundamental truths, like mathematics. Within the IT industry, and especially in B2B contexts, the logic of universality has also been useful for driving scalability; a technology that is framed and assumed to be easily repurposed and sold to multiple clients is generally understood as good for business. However, when approached through the lens of sociotechnical systems, universality breaks down, as there are social, cultural, religious, economic, political, and epistemological factors that must be accounted for and that are unique to specific places, time periods, social groups, or communities~\cite{Selbst2019}. In the example of the patients in the vignette above, universality breaks down as soon as we consider the contextual specificity of the lived realities of the patients--some of whom may not have access to the broader infrastructures or skilling needed to interact with the technology, or for whom the technology may not even work due to geographic, cultural, or embodied differences. 

Finally, we call attention to \textbf{techno-beneficence}, which we define as \textit{the presumption that all technological developments and insights inherently contribute to the betterment of society}. Within techno-beneficence, technology-mediated harms are understood as either unintended consequences or as caused by nefarious actors. The underlying logic behind such causal claims is that the creation of technology or the production of insights itself was good, but then (i.e., afterwards) something bad happened. This presumption of a priori beneficence is deep rooted in most forms of institutionalized technology development and scientific practice because funding and approval for projects typically demand an articulation of projects as inherently pursuit-worthy. In the B2B context, there is an added incentive to frame projects as good for business. The problem with techno-beneficence as a \textit{systematic assumption}, however, is that it precludes discussions on whether something should be built in the first place. It also enables the compartmentalization of social responsibility, framing technology creators' role as that of producing a well-performing product, without the need to engage in activities that ensure the project is in fact beneficial for society. 

The goal of outlining and being cognizant of these common assumptions within the R\&I Framework is not to say that technology is never the solution, that it can never be repurposed, nor that it can never be beneficial. Rather, the goal is to re-framed these assumptions as a \textit{social hypothesis} that, like any other hypothesis, needs to be tested and grounded in not just quantitative but also qualitative evidence about the specific socio-historical contexts in which technology will be used.

\subsubsection{Bridging the gap between contexts of creation and contexts of use} 
 
Implicitly and explicitly, throughout the R\&I Framework we situate critical reflection as a core component of \textbf{responsible technology practice} because of its role in bridging the accountability and awareness gaps. We recognize that robust accountability within complex business hierarchies or well-entrenched corporate logics will require more than reflexivity. However, drawing attention to societal stakeholders, social impacts, and socio-historical considerations has the potential to foster a culture in which technical, governance, and business decision-makers ``grow accustomed to critically engaging and reformulating the assumptions and epistemological leaps under-girding the research [and business] opportunities they are offered''~\cite{Barabas2020}. Through this reflexivity practitioners may find that they are more morally entangled than they previously thought, which in turn could lead to more efforts to establish connectivity across the technology supply chain. 

Through its conceptual framing and structure, the R\&I Framework also nudges towards \textbf{inclusive technology practice}. Although inclusion is fraught, especially given efforts to extract value and extend power through predatory inclusion\cite{Hoffman2021}, reflexive engagements with societal stakeholders and collaborations with people that are critical of normative ways of knowing can expand the scope of mitigative actions. This could involve increasing the representation of people with diverse backgrounds and identities, especially groups traditionally underrepresented in Science, Technology, and Engineering (STEM). It could also entail putting together cross-disciplinary teams so that multiple approaches, methods, and types of insights may be brought to bear early on within a project. More situated engagements with communities to include their perspectives, needs, and concerns could also be realized through participatory methods or co-design. Additionally, procedures that forge connections between the contexts of technology creation and use could be set in place, including but not limited to establishing feedback mechanisms, acceptable use policies, and thorough impact assessments. 

The goal of these various mitigative actions should be to ``reveal not only the technical features and limitations of a system, but also the different ways a system affects different people in its context of use.'' \cite{katell2020}. As such, the success of these actions hinges on participants' and stakeholders' ability to do more than validate or check-mark a project. This requires giving self, business, and society stakeholders the opportunity to set the terms of their engagement and have decision-making power throughout the project, including refusing to partake or suggesting a project should be stopped because it is unethical or harmful. We argue that balancing these commitments with business and technical considerations, which are critical to the functioning of B2B companies, is not only possible but necessary to prevent and mitigate technology-mediated harms \cite{Schiff2020}.


\subsection{Applying the R\&I Framework}
 
Applying the R\&I Framework once at the beginning of a project, a client engagement, or ethical review will not in and of itself mitigate harms. First, to drive cultural change, the process of critically interrogating stakeholders, incentives, and socio-historical contexts ought to become routinary--it needs to become part of governance mechanisms, organizational practices, and incentive structures. This is in line with Selbst et al.'s notion of ``constructive reform'' \cite{Selbst2019}, particularly the re-orientation of technology work away from solutions and towards socially-grounded processes, and Leidner at al.'s framework for culture change \cite{Leidner2006}, on exposing and reconciling conflicting values. 

Second, to mitigate harms, cultural change must be accompanied by actions. These mitigative actions could range from adapting a product based on participatory workshops with societal stakeholders, all the way to deciding that a particular technology should not be built or sold. In its current form, the R\&I Framework does not offer guidance on which course of action to follow; however, it does offer a way of more comprehensively understanding \textit{why} harms need to be mitigated, \textit{what} kind of mitigation strategies are possible, and \textit{who} stands to benefit or not from a given mitigation strategy. In this way, the R\&I Framework can be applied at any stage of the technology life-cycle to broaden the scope of what is considered possible in terms of advancing fairness, accountability, and transparency in and through sociotechnical systems. 

For people interested in collaboratively engaging with aspects of the framework, we have designed an initial set of tools including a  mural\footnote{http://app.mural.co/} map for identifying stakeholders and impacts\footnote{
https://app.mural.co/template/d6f4d823-4fa4-4662-bc3e-9c48425cc322/f2162414-778a-49ba-83d6-e2ce4376e6b9
}, an interactive card tool for promoting critical thinking for teams in a turn-taking democratized way\footnote{
https://incltech.mybluemix.net/incltech/static/game.html
}, and a chatbot to to promote critical thinking around impact statements. 
The mural map depicts self, business, and society forces in a way that supports the identification of stakeholders (in the expanded notion proposed by the framework) and indirect-delayed impacts, under-considered harms, and opportunities. The interactive card tool has a set of 107 questions, structured in different levels of depth, to guide and promote critical reflection and group discussions covering multiple elements of the framework such as unpacking socio-historical context, stakeholders mapping, reflecting about impact \& outcomes, and sharing practices \& actions. The card tool includes questions such as \textit{"Are there any stakeholders who might not know they could be impacted by your work?" }or \textit{"Of the stakeholders with the least decision-making power, who would be the most impacted by your work?"} Beyond these tools supporting design and development stages, our team also interacts with other teams in our organization to incorporate aspects from the R\&I Framework as part of existing processes and practices ranging from design processes to project management.
The existing artifacts and processes interventions represent our first instantiations of the R\&I Framework and do not exhaust all possibilities for technologists and practitioners to be responsive to existing needs.

We recognize that it is not easy or sometimes even possible to cede power \cite{EXD}, especially since the people designing or building the technologies in B2B companies are usually not the same as those applying them and may not have much influence beyond contexts of creations; they may have little if any say in how the technologies are used or who they are sold to, for example.  However, our hope is that by revealing differential impacts and considerations within specific projects or larger sociotechnological processes and calling attention to common assumptions and narratives that are problematic or limiting, we might help technologists and practitioners find more \textit{subtle} ways to confront power in their daily practice. We invite and welcome other interpretations of the framework, as well as its transformation into practices, strategies, conceptualizations, tools, and methodological approaches.
\section{Limitations and Open Questions}
\label{limitations}

There are a number of limitations to implementing the framework, some which directly reflect the tensions and challenges outlined in the paper. One limitation is that corporate structures and incentives may limit how much time people have to engage in critical reflection and the extent to which they can prioritize societal considerations over other considerations (e.g., financial gains and technical demands). Related to this, the opacity of business practices makes it so that stakeholders may not know all incentives, motivations, and forces that impinge on a project, thus restricting their ability to interrogate practices and decisions. Another limitation is the already widespread belief that responsibility is limited to compliance, as well as the common practice of measuring the success of responsible and inclusive practices based on reputational metrics, rather than on-the-ground observations of social impacts.

These limitations are compounded by our restricted ability to demonstrate how the framework could be implemented, as well as how to resolve barriers that emerged during implementation. Our engagements with researchers, developers, designers, and consultants involved proprietary projects, which left us unable to detail these specific examples. We also have yet to explore the applicability of the framework's concepts to other contexts 
and understand if this way of framing responsibility and inclusion translates to other audiences, e.g., people in academic settings, direct to consumer (DTC) or business to consumer (B2C) models, and smaller companies. We thus recognize that there may be crucial considerations missing within our framework that would be critical to the effective implementation of the framework in a variety of settings.

These challenges lead to important but open questions for future work.  The main question that continuously resurfaces for our team is, \textbf{"How do we measure the impact of the framework?"}  Part of the difficulty in measuring impact has to do with the complexity of the environment and the nuance of the issues. Additionally, the size of the company and the B2B model makes it hard to address the principles/theory-to-practice gap, particularly when it comes to larger system-wide impacts or impacts that occur beyond the scope of a client contract.  For example, one of the goals of our initiative is to positively impact the company's culture by changing technology practices - but we are still determining how to show cultural change has occurred and the associated scale. 
We welcome ideas for thinking through the assessment and measurement of impact in ways that are simultaneously critical of over-simplification and reductionism while also attentive to the aim to evaluate value. 

\section{Conclusion}
\label{conclusion}

In this paper, we proposed the R\&I Framework, a formative framework to orient critical reflection around the social contexts of technology creation and use; the power dynamics between self, business, and societal stakeholders; historical considerations relating to the impacts of technology on various communities; 
and the practical decisions that imbue technological artifacts with cultural values.
After presenting existing frameworks, principles, and critical reflection tools, we detailed how the B2B context may not be suitably covered by existing approaches. 
Key factors brought by the proposed framework in terms of \textbf{the politics of technology creation and use} and \textbf{the socio-historical contexts of technology creation and use} have the potential to shorten the distance between the creation of technology and indirect-delayed impacts and call attention to existing narratives, which may amplify 'responsible blind spots' in the IT industry.
Finally, B2B contexts have an inherent potential to dilute responsibility, to be less inclusive, and, hence, to have an even more challenging responsibility gap to solve. Our framework connects these aspects, 
shows a direction we are 
following, and offers opportunities for 
others to reuse and expand.

\begin{acks}
We thank 
Adriana Alvarado Garcia, Amber Hamilton, Aminat Adebiyi, Alex Baria, Felicia Jing, Mira Luca Wolf-Bauwens, Rogerio Abreu de Paula, and Salma Elsayed-Ali
for all thoughtful discussions and valuable feedback during our interactions.

\end{acks}

\bibliographystyle{ACM-Reference-Format}
\bibliography{0-Main}


\begin{thebibliography}{99}


\ifx \showCODEN    \undefined \def \showCODEN     #1{\unskip}     \fi
\ifx \showDOI      \undefined \def \showDOI       #1{#1}\fi
\ifx \showISBNx    \undefined \def \showISBNx     #1{\unskip}     \fi
\ifx \showISBNxiii \undefined \def \showISBNxiii  #1{\unskip}     \fi
\ifx \showISSN     \undefined \def \showISSN      #1{\unskip}     \fi
\ifx \showLCCN     \undefined \def \showLCCN      #1{\unskip}     \fi
\ifx \shownote     \undefined \def \shownote      #1{#1}          \fi
\ifx \showarticletitle \undefined \def \showarticletitle #1{#1}   \fi
\ifx \showURL      \undefined \def \showURL       {\relax}        \fi
\providecommand\bibfield[2]{#2}
\providecommand\bibinfo[2]{#2}
\providecommand\natexlab[1]{#1}
\providecommand\showeprint[2][]{arXiv:#2}

\bibitem[Abbate and Dick(2022)]%
        {Abbate2022}
\bibfield{editor}{\bibinfo{person}{Janet Abbate} {and}
  \bibinfo{person}{Stephanie Dick}} (Eds.). \bibinfo{year}{2022}\natexlab{}.
\newblock \bibinfo{booktitle}{\emph{Abstractions and Embodiments: New Histories
  of Computing and Society}}.
\newblock \bibinfo{publisher}{John Hopkins University Press}.
\newblock


\bibitem[Abu El-Haj(2002)]%
        {AbuElHaj2002}
\bibfield{author}{\bibinfo{person}{Nadia Abu El-Haj}.}
  \bibinfo{year}{2002}\natexlab{}.
\newblock \bibinfo{booktitle}{\emph{Facts on the Ground: Archeological Practice
  and Territorial Self-Fashioning in Israeli Society}}.
\newblock \bibinfo{publisher}{The University of Chicago Press}.
\newblock


\bibitem[Agafonow(2017)]%
        {Agafonow}
\bibfield{author}{\bibinfo{person}{Alejandro Agafonow}.}
  \bibinfo{year}{2017}\natexlab{}.
\newblock \showarticletitle{Design Thinking and Social Enterprises: A
  Solution-Focused Strategy for Social Enterprise Research}.
\newblock \bibinfo{journal}{\emph{European Management Review}}
  (\bibinfo{year}{2017}).
\newblock
\urldef\tempurl%
\url{https://doi.org/10.1111/emre.12152}
\showDOI{\tempurl}


\bibitem[Anderson(2006)]%
        {Anderson2006}
\bibfield{author}{\bibinfo{person}{Warwick Anderson}.}
  \bibinfo{year}{2006}\natexlab{}.
\newblock \bibinfo{booktitle}{\emph{Colonial Pathologies: American Tropical
  Medicine, Race, and Hygiene in the Philippines}}.
\newblock \bibinfo{publisher}{Duke University Press}.
\newblock


\bibitem[Arnold et~al\mbox{.}(2019)]%
        {Arnold2019}
\bibfield{author}{\bibinfo{person}{M. Arnold}, \bibinfo{person}{R.~K.~E.
  Bellamy}, \bibinfo{person}{M. Hind}, \bibinfo{person}{S. Houde},
  \bibinfo{person}{S. Mehta}, \bibinfo{person}{A. Mojsilović},
  \bibinfo{person}{R. Nair}, \bibinfo{person}{K.~Natesan Ramamurthy},
  \bibinfo{person}{A. Olteanu}, \bibinfo{person}{D. Piorkowski},
  \bibinfo{person}{D. Reimer}, \bibinfo{person}{J. Richards},
  \bibinfo{person}{J. Tsay}, {and} \bibinfo{person}{K.~R. Varshney}.}
  \bibinfo{year}{2019}\natexlab{}.
\newblock \showarticletitle{FactSheets: Increasing trust in AI services through
  supplier's declarations of conformity}.
\newblock \bibinfo{journal}{\emph{IBM Journal of Research and Development}}
  \bibinfo{volume}{63}, \bibinfo{number}{4/5} (\bibinfo{year}{2019}),
  \bibinfo{pages}{6:1--6:13}.
\newblock
\urldef\tempurl%
\url{https://doi.org/10.1147/JRD.2019.2942288}
\showDOI{\tempurl}


\bibitem[Atanasoski and Vora(2019)]%
        {Atanasoski2019}
\bibfield{author}{\bibinfo{person}{Neda Atanasoski} {and}
  \bibinfo{person}{Kalindi Vora}.} \bibinfo{year}{2019}\natexlab{}.
\newblock \bibinfo{booktitle}{\emph{Surrogate Humanity: Race, Robots, and the
  Politics of Technological Futures}}.
\newblock \bibinfo{publisher}{Duke University Press}.
\newblock


\bibitem[Ballard et~al\mbox{.}(2019)]%
        {Ballard2019}
\bibfield{author}{\bibinfo{person}{Stephanie Ballard}, \bibinfo{person}{Karen~M
  Chappell}, {and} \bibinfo{person}{Kristen Kennedy}.}
  \bibinfo{year}{2019}\natexlab{}.
\newblock \showarticletitle{Judgment call the game: Using value sensitive
  design and design fiction to surface ethical concerns related to technology}.
  In \bibinfo{booktitle}{\emph{Proceedings of the 2019 on Designing Interactive
  Systems Conference}}. \bibinfo{pages}{421--433}.
\newblock


\bibitem[Barabas et~al\mbox{.}(2020)]%
        {Barabas2020}
\bibfield{author}{\bibinfo{person}{Chelsea Barabas}, \bibinfo{person}{Colin
  Doyle}, \bibinfo{person}{JB Rubinovitz}, {and} \bibinfo{person}{Karthik
  Dinakar}.} \bibinfo{year}{2020}\natexlab{}.
\newblock \showarticletitle{Studying up: Reorienting the Study of Algorithmic
  Fairness around Issues of Power}. In \bibinfo{booktitle}{\emph{Proceedings of
  the 2020 Conference on Fairness, Accountability, and Transparency}}
  (Barcelona, Spain) \emph{(\bibinfo{series}{FAT* '20})}.
  \bibinfo{publisher}{Association for Computing Machinery},
  \bibinfo{address}{New York, NY, USA}, \bibinfo{pages}{167–176}.
\newblock
\showISBNx{9781450369367}
\urldef\tempurl%
\url{https://doi.org/10.1145/3351095.3372859}
\showDOI{\tempurl}


\bibitem[Benjamin(2019)]%
        {Benjamin2019}
\bibfield{author}{\bibinfo{person}{Ruha Benjamin}.}
  \bibinfo{year}{2019}\natexlab{}.
\newblock \bibinfo{booktitle}{\emph{Race After Technology: Abolitionist Tools
  for the New Jim Code}}.
\newblock \bibinfo{publisher}{Polity}.
\newblock


\bibitem[Benjamins(2021)]%
        {Benjamins2021}
\bibfield{author}{\bibinfo{person}{Richard Benjamins}.}
  \bibinfo{year}{2021}\natexlab{}.
\newblock \showarticletitle{A choices framework for the responsible use of AI}.
\newblock \bibinfo{journal}{\emph{AI and Ethics}} \bibinfo{volume}{1},
  \bibinfo{number}{1} (\bibinfo{year}{2021}), \bibinfo{pages}{49--53}.
\newblock


\bibitem[Bietti(2020)]%
        {Bietti2020}
\bibfield{author}{\bibinfo{person}{Elettra Bietti}.}
  \bibinfo{year}{2020}\natexlab{}.
\newblock \showarticletitle{From ethics washing to ethics bashing: a view on
  tech ethics from within moral philosophy}. In
  \bibinfo{booktitle}{\emph{Proceedings of the 2020 conference on fairness,
  accountability, and transparency}}. \bibinfo{pages}{210--219}.
\newblock


\bibitem[Bj{\"o}rgvinsson et~al\mbox{.}(2012)]%
        {Bjorgvinsson2012-kp}
\bibfield{author}{\bibinfo{person}{Erling Bj{\"o}rgvinsson},
  \bibinfo{person}{Pelle Ehn}, {and} \bibinfo{person}{Per-Anders Hillgren}.}
  \bibinfo{year}{2012}\natexlab{}.
\newblock \showarticletitle{Agonistic participatory design: working with
  marginalised social movements}.
\newblock \bibinfo{journal}{\emph{CoDesign}} \bibinfo{volume}{8},
  \bibinfo{number}{2-3} (\bibinfo{date}{June} \bibinfo{year}{2012}),
  \bibinfo{pages}{127--144}.
\newblock


\bibitem[Buhmann and Fieseler(2021)]%
        {Buhmann2021}
\bibfield{author}{\bibinfo{person}{Alexander Buhmann} {and}
  \bibinfo{person}{Christian Fieseler}.} \bibinfo{year}{2021}\natexlab{}.
\newblock \showarticletitle{Towards a deliberative framework for responsible
  innovation in artificial intelligence}.
\newblock \bibinfo{journal}{\emph{Technology in Society}}  \bibinfo{volume}{64}
  (\bibinfo{year}{2021}), \bibinfo{pages}{101475}.
\newblock
\showISSN{0160-791X}
\urldef\tempurl%
\url{https://doi.org/10.1016/j.techsoc.2020.101475}
\showDOI{\tempurl}


\bibitem[Calderon~Salazar and Huybrechts(2020)]%
        {Calderon_Salazar2020-cm}
\bibfield{author}{\bibinfo{person}{Pablo Calderon~Salazar} {and}
  \bibinfo{person}{Liesbeth Huybrechts}.} \bibinfo{year}{2020}\natexlab{}.
\newblock \showarticletitle{{PD} otherwise will be pluriversal (or it won't
  be)}. In \bibinfo{booktitle}{\emph{Proceedings of the 16th Participatory
  Design Conference 2020 - Participation(s) Otherwise - Volume 1}} (Manizales
  Colombia). \bibinfo{publisher}{ACM}, \bibinfo{address}{New York, NY, USA}.
\newblock


\bibitem[Charlotte~Smith et~al\mbox{.}(2020)]%
        {Charlotte_Smith2020-zg}
\bibfield{author}{\bibinfo{person}{Rachel Charlotte~Smith},
  \bibinfo{person}{Heike Winschiers-Theophilus}, \bibinfo{person}{Daria Loi},
  \bibinfo{person}{Asnath Paula~Kambunga}, \bibinfo{person}{Marly
  Muudeni~Samuel}, {and} \bibinfo{person}{Rogerio de Paula}.}
  \bibinfo{year}{2020}\natexlab{}.
\newblock \showarticletitle{Decolonising Participatory Design Practices:
  Towards Participations Otherwise}. In \bibinfo{booktitle}{\emph{Proceedings
  of the 16th Participatory Design Conference 2020 - Participation(s) Otherwise
  - Volume 2}} \emph{(\bibinfo{series}{PDC '20})}.
  \bibinfo{publisher}{Association for Computing Machinery},
  \bibinfo{address}{New York, NY, USA}, \bibinfo{pages}{206--208}.
\newblock


\bibitem[Cipolla et~al\mbox{.}(2017)]%
        {Cipolla2017}
\bibfield{editor}{\bibinfo{person}{Cyd Cipolla}, \bibinfo{person}{Kristina
  Gupta}, \bibinfo{person}{David Rubin}, {and} \bibinfo{person}{Angela Willey}}
  (Eds.). \bibinfo{year}{2017}\natexlab{}.
\newblock \bibinfo{booktitle}{\emph{Queer Feminist Science Studies: A Reader}}.
\newblock \bibinfo{publisher}{University of Washington Press}.
\newblock


\bibitem[Collins(1986)]%
        {Collins1986}
\bibfield{author}{\bibinfo{person}{Patricia~Hill Collins}.}
  \bibinfo{year}{1986}\natexlab{}.
\newblock \showarticletitle{Learning from the outsider within: The sociological
  significance of Black feminist thought}.
\newblock \bibinfo{journal}{\emph{Social problems}} \bibinfo{volume}{33},
  \bibinfo{number}{6} (\bibinfo{year}{1986}), \bibinfo{pages}{s14--s32}.
\newblock


\bibitem[Contractor et~al\mbox{.}(2022)]%
        {Contractor2022}
\bibfield{author}{\bibinfo{person}{Danish Contractor}, \bibinfo{person}{Daniel
  McDuff}, \bibinfo{person}{Julia~Katherine Haines}, \bibinfo{person}{Jenny
  Lee}, \bibinfo{person}{Christopher Hines}, \bibinfo{person}{Brent Hecht},
  \bibinfo{person}{Nicholas Vincent}, {and} \bibinfo{person}{Hanlin Li}.}
  \bibinfo{year}{2022}\natexlab{}.
\newblock \showarticletitle{Behavioral use licensing for responsible AI}. In
  \bibinfo{booktitle}{\emph{2022 ACM Conference on Fairness, Accountability,
  and Transparency}}. \bibinfo{pages}{778--788}.
\newblock


\bibitem[Crawford(2021)]%
        {Crawford2021}
\bibfield{author}{\bibinfo{person}{Kate Crawford}.}
  \bibinfo{year}{2021}\natexlab{}.
\newblock \bibinfo{booktitle}{\emph{The atlas of AI: Power, politics, and the
  planetary costs of artificial intelligence}}.
\newblock \bibinfo{publisher}{Yale University Press}.
\newblock


\bibitem[Daston and Galison(2007)]%
        {Daston2007}
\bibfield{author}{\bibinfo{person}{Lorraine Daston} {and}
  \bibinfo{person}{Peter Galison}.} \bibinfo{year}{2007}\natexlab{}.
\newblock \bibinfo{booktitle}{\emph{Objectivity}}.
\newblock \bibinfo{publisher}{Princeton University Press}.
\newblock


\bibitem[David et~al\mbox{.}(2021)]%
        {Hendry2021}
\bibfield{author}{\bibinfo{person}{Hendry David}, \bibinfo{person}{Batya
  Friedman}, {and} \bibinfo{person}{Stephanie Ballard}.}
  \bibinfo{year}{2021}\natexlab{}.
\newblock \showarticletitle{Value Sensitive Design as a Formative Framework}.
\newblock \bibinfo{journal}{\emph{Ethics and Information Technology}}
  \bibinfo{volume}{23} (\bibinfo{year}{2021}), \bibinfo{pages}{39--44}.
\newblock


\bibitem[D\'{\i}az et~al\mbox{.}(2022)]%
        {Diaz2022}
\bibfield{author}{\bibinfo{person}{Mark D\'{\i}az}, \bibinfo{person}{Ian
  Kivlichan}, \bibinfo{person}{Rachel Rosen}, \bibinfo{person}{Dylan Baker},
  \bibinfo{person}{Razvan Amironesei}, \bibinfo{person}{Vinodkumar
  Prabhakaran}, {and} \bibinfo{person}{Emily Denton}.}
  \bibinfo{year}{2022}\natexlab{}.
\newblock \showarticletitle{CrowdWorkSheets: Accounting for Individual and
  Collective Identities Underlying Crowdsourced Dataset Annotation}. In
  \bibinfo{booktitle}{\emph{2022 ACM Conference on Fairness, Accountability,
  and Transparency}} (Seoul, Republic of Korea) \emph{(\bibinfo{series}{FAccT
  '22})}. \bibinfo{publisher}{Association for Computing Machinery},
  \bibinfo{address}{New York, NY, USA}, \bibinfo{pages}{2342–2351}.
\newblock
\showISBNx{9781450393522}
\urldef\tempurl%
\url{https://doi.org/10.1145/3531146.3534647}
\showDOI{\tempurl}


\bibitem[D'Ignazio et~al\mbox{.}(2020)]%
        {DIgnazio2020-yj}
\bibfield{author}{\bibinfo{person}{Catherine D'Ignazio},
  \bibinfo{person}{Erhardt Graeff}, \bibinfo{person}{Christina~N Harrington},
  {and} \bibinfo{person}{Daniela~K Rosner}.} \bibinfo{year}{2020}\natexlab{}.
\newblock \showarticletitle{Toward Equitable Participatory Design: Data
  Feminism for {CSCW} amidst Multiple Pandemics}. In
  \bibinfo{booktitle}{\emph{Conference Companion Publication of the 2020 on
  Computer Supported Cooperative Work and Social Computing}} (Virtual Event,
  USA) \emph{(\bibinfo{series}{CSCW '20 Companion})}.
  \bibinfo{publisher}{Association for Computing Machinery},
  \bibinfo{address}{New York, NY, USA}, \bibinfo{pages}{437--445}.
\newblock


\bibitem[D'ignazio and Klein(2020)]%
        {Dignazio2020}
\bibfield{author}{\bibinfo{person}{Catherine D'ignazio} {and}
  \bibinfo{person}{Lauren~F Klein}.} \bibinfo{year}{2020}\natexlab{}.
\newblock \bibinfo{booktitle}{\emph{Data feminism}}.
\newblock \bibinfo{publisher}{MIT press}.
\newblock


\bibitem[Dunne and Raby(2013)]%
        {Dunne2013}
\bibfield{author}{\bibinfo{person}{Anthony Dunne} {and} \bibinfo{person}{Fiona
  Raby}.} \bibinfo{year}{2013}\natexlab{}.
\newblock \bibinfo{booktitle}{\emph{Speculative everything: design, fiction,
  and social dreaming}}.
\newblock \bibinfo{publisher}{MIT press}.
\newblock


\bibitem[Flyvbjerg et~al\mbox{.}(2012)]%
        {Flyvbjerg2012}
\bibfield{author}{\bibinfo{person}{B Flyvbjerg}, \bibinfo{person}{T Landman},
  {and} \bibinfo{person}{Schram S}.} \bibinfo{year}{2012}\natexlab{}.
\newblock \bibinfo{booktitle}{\emph{Real Social Science: Applied Phronesis}}.
\newblock \bibinfo{publisher}{Cambridge University Press}.
\newblock


\bibitem[Foth and Axup(2006)]%
        {Foth2006}
\bibfield{author}{\bibinfo{person}{Marcus Foth} {and} \bibinfo{person}{Jeff
  Axup}.} \bibinfo{year}{2006}\natexlab{}.
\newblock \showarticletitle{Participatory Design and Action Research: Identical
  Twins or Synergetic Pair?}. In \bibinfo{booktitle}{\emph{Expanding Boundaries
  in Design: Proceedings Ninth Participatory Design Conference}} (Canada),
  Vol.~\bibinfo{volume}{2}. \bibinfo{publisher}{Computer Professionals for
  Social Responsibility}, \bibinfo{pages}{93--96}.
\newblock


\bibitem[Freeman and Gilbert(1992)]%
        {Freeman1992}
\bibfield{author}{\bibinfo{person}{R.E Freeman} {and} \bibinfo{person}{D.R.
  Gilbert}.} \bibinfo{year}{1992}\natexlab{}.
\newblock \showarticletitle{Business, Ethics and Society: A Critical Agenda}.
\newblock \bibinfo{journal}{\emph{Business and Society}} \bibinfo{volume}{31},
  \bibinfo{number}{1} (\bibinfo{year}{1992}), \bibinfo{pages}{9--17}.
\newblock


\bibitem[Friedman(1996)]%
        {Friedman1996}
\bibfield{author}{\bibinfo{person}{Batya Friedman}.}
  \bibinfo{year}{1996}\natexlab{}.
\newblock \showarticletitle{Value-sensitive design}.
\newblock \bibinfo{journal}{\emph{Interactions}} \bibinfo{volume}{3},
  \bibinfo{number}{6} (\bibinfo{year}{1996}), \bibinfo{pages}{16--23}.
\newblock


\bibitem[Friedman and Hendry(2012)]%
        {Friedman2012}
\bibfield{author}{\bibinfo{person}{Batya Friedman} {and} \bibinfo{person}{David
  Hendry}.} \bibinfo{year}{2012}\natexlab{}.
\newblock \showarticletitle{The envisioning cards: a toolkit for catalyzing
  humanistic and technical imaginations}. In
  \bibinfo{booktitle}{\emph{Proceedings of the SIGCHI conference on human
  factors in computing systems}}. \bibinfo{pages}{1145--1148}.
\newblock


\bibitem[Fullwiley(2012)]%
        {Fullwiley2012}
\bibfield{author}{\bibinfo{person}{Duana Fullwiley}.}
  \bibinfo{year}{2012}\natexlab{}.
\newblock \bibinfo{booktitle}{\emph{The Enculturated Gene: Sickle Cell Health
  Politics and Biological Difference in West Africa}}.
\newblock \bibinfo{publisher}{Princeton University Press}.
\newblock


\bibitem[Gansky and McDonald(2022)]%
        {Gansky2022}
\bibfield{author}{\bibinfo{person}{Ben Gansky} {and} \bibinfo{person}{Sean
  McDonald}.} \bibinfo{year}{2022}\natexlab{}.
\newblock \showarticletitle{CounterFAccTual: How FAccT Undermines Its
  Organizing Principles}. In \bibinfo{booktitle}{\emph{2022 ACM Conference on
  Fairness, Accountability, and Transparency}} (Seoul, Republic of Korea)
  \emph{(\bibinfo{series}{FAccT '22})}. \bibinfo{publisher}{Association for
  Computing Machinery}, \bibinfo{address}{New York, NY, USA},
  \bibinfo{pages}{1982–1992}.
\newblock
\showISBNx{9781450393522}
\urldef\tempurl%
\url{https://doi.org/10.1145/3531146.3533241}
\showDOI{\tempurl}


\bibitem[Globus-Harris et~al\mbox{.}(2022)]%
        {Globus-Harris2022}
\bibfield{author}{\bibinfo{person}{Ira Globus-Harris}, \bibinfo{person}{Michael
  Kearns}, {and} \bibinfo{person}{Aaron Roth}.}
  \bibinfo{year}{2022}\natexlab{}.
\newblock \showarticletitle{An Algorithmic Framework for Bias Bounties}. In
  \bibinfo{booktitle}{\emph{2022 ACM Conference on Fairness, Accountability,
  and Transparency}} (Seoul, Republic of Korea) \emph{(\bibinfo{series}{FAccT
  '22})}. \bibinfo{publisher}{Association for Computing Machinery},
  \bibinfo{address}{New York, NY, USA}, \bibinfo{pages}{1106–1124}.
\newblock
\showISBNx{9781450393522}
\urldef\tempurl%
\url{https://doi.org/10.1145/3531146.3533172}
\showDOI{\tempurl}


\bibitem[Goetze(2021)]%
        {Goetze2021}
\bibfield{author}{\bibinfo{person}{Trystan~S Goetze}.}
  \bibinfo{year}{2021}\natexlab{}.
\newblock \showarticletitle{Moral Entanglement: Taking Responsibility and
  Vicarious Responsibility}.
\newblock \bibinfo{journal}{\emph{The Monist}} \bibinfo{volume}{104},
  \bibinfo{number}{2} (\bibinfo{year}{2021}), \bibinfo{pages}{210--223}.
\newblock


\bibitem[Goetze(2022)]%
        {Goetze2022}
\bibfield{author}{\bibinfo{person}{Trystan~S. Goetze}.}
  \bibinfo{year}{2022}\natexlab{}.
\newblock \showarticletitle{Mind the Gap: Autonomous Systems, the
  Responsibility Gap, and Moral Entanglement}. In
  \bibinfo{booktitle}{\emph{2022 ACM Conference on Fairness, Accountability,
  and Transparency}} (Seoul, Republic of Korea) \emph{(\bibinfo{series}{FAccT
  '22})}. \bibinfo{publisher}{Association for Computing Machinery},
  \bibinfo{address}{New York, NY, USA}, \bibinfo{pages}{390–400}.
\newblock
\showISBNx{9781450393522}
\urldef\tempurl%
\url{https://doi.org/10.1145/3531146.3533106}
\showDOI{\tempurl}


\bibitem[Green(2020)]%
        {Green2020}
\bibfield{author}{\bibinfo{person}{Ben Green}.}
  \bibinfo{year}{2020}\natexlab{}.
\newblock \showarticletitle{The False Promise of Risk Assessments: Epistemic
  Reform and the Limits of Fairness}. In \bibinfo{booktitle}{\emph{Proceedings
  of the 2020 Conference on Fairness, Accountability, and Transparency}}
  (Barcelona, Spain) \emph{(\bibinfo{series}{FAT* '20})}.
  \bibinfo{publisher}{Association for Computing Machinery},
  \bibinfo{address}{New York, NY, USA}, \bibinfo{pages}{594–606}.
\newblock
\showISBNx{9781450369367}
\urldef\tempurl%
\url{https://doi.org/10.1145/3351095.3372869}
\showDOI{\tempurl}


\bibitem[Greene et~al\mbox{.}(2019)]%
        {Greene2019}
\bibfield{author}{\bibinfo{person}{Daniel Greene}, \bibinfo{person}{Anna~Lauren
  Hoffmann}, {and} \bibinfo{person}{Luke Stark}.}
  \bibinfo{year}{2019}\natexlab{}.
\newblock \showarticletitle{Better, nicer, clearer, fairer: A critical
  assessment of the movement for ethical artificial intelligence and machine
  learning}.
\newblock  (\bibinfo{year}{2019}).
\newblock


\bibitem[Hammonds and Herzig(2009)]%
        {Hammonds2009}
\bibfield{author}{\bibinfo{person}{Evelynn~M. Hammonds} {and}
  \bibinfo{person}{Rebecca~M. Herzig}.} \bibinfo{year}{2009}\natexlab{}.
\newblock \bibinfo{booktitle}{\emph{The Nature of Difference: Sciences of Race
  in the United States from Jefferson to Genomics}}.
\newblock \bibinfo{publisher}{The MIT Press}.
\newblock


\bibitem[Haraway(1990)]%
        {Haraway1990}
\bibfield{author}{\bibinfo{person}{Donna Haraway}.}
  \bibinfo{year}{1990}\natexlab{}.
\newblock \bibinfo{booktitle}{\emph{Simians, Cyborgs, and Women: The
  Reinvention of Nature}}.
\newblock \bibinfo{publisher}{Routledge}.
\newblock


\bibitem[Harding(2003)]%
        {Harding2003}
\bibfield{editor}{\bibinfo{person}{Sandra Harding}} (Ed.).
  \bibinfo{year}{2003}\natexlab{}.
\newblock \bibinfo{booktitle}{\emph{The Feminist Standpoint Theory Reader:
  Intellectual and Political Controversies}}.
\newblock \bibinfo{publisher}{Routledge}.
\newblock


\bibitem[Harrington and Dillahunt(2021)]%
        {Harrington2021}
\bibfield{author}{\bibinfo{person}{Christina Harrington} {and}
  \bibinfo{person}{Tawanna~R Dillahunt}.} \bibinfo{year}{2021}\natexlab{}.
\newblock \showarticletitle{Eliciting Tech Futures Among Black Young Adults: A
  Case Study of Remote Speculative Co-Design}. In
  \bibinfo{booktitle}{\emph{Proceedings of the 2021 CHI Conference on Human
  Factors in Computing Systems}} (Yokohama, Japan) \emph{(\bibinfo{series}{CHI
  '21})}. \bibinfo{publisher}{Association for Computing Machinery},
  \bibinfo{address}{New York, NY, USA}, Article \bibinfo{articleno}{397},
  \bibinfo{numpages}{15}~pages.
\newblock
\showISBNx{9781450380966}
\urldef\tempurl%
\url{https://doi.org/10.1145/3411764.3445723}
\showDOI{\tempurl}


\bibitem[Harrington et~al\mbox{.}(2019)]%
        {Harrington2019-pv}
\bibfield{author}{\bibinfo{person}{Christina Harrington},
  \bibinfo{person}{Sheena Erete}, {and} \bibinfo{person}{Anne~Marie Piper}.}
  \bibinfo{year}{2019}\natexlab{}.
\newblock \showarticletitle{Deconstructing {Community-Based} Collaborative
  Design: Towards More Equitable Participatory Design Engagements}.
\newblock \bibinfo{journal}{\emph{Proc. ACM Hum.-Comput. Interact.}}
  \bibinfo{volume}{3}, \bibinfo{number}{CSCW} (\bibinfo{date}{Nov.}
  \bibinfo{year}{2019}), \bibinfo{pages}{1--25}.
\newblock


\bibitem[Hecht(2014)]%
        {Hecht2014}
\bibfield{author}{\bibinfo{person}{Gabrielle Hecht}.}
  \bibinfo{year}{2014}\natexlab{}.
\newblock \bibinfo{booktitle}{\emph{Being Nuclear: Africans and the Global
  Uranium Trade}}.
\newblock \bibinfo{publisher}{The MIT Press}.
\newblock


\bibitem[Hill et~al\mbox{.}(2016)]%
        {EXD}
\bibfield{author}{\bibinfo{person}{Caroline Hill}, \bibinfo{person}{Michelle
  Molitor}, {and} \bibinfo{person}{Christine Ortiz}.}
  \bibinfo{year}{2016}\natexlab{}.
\newblock \bibinfo{title}{Racism and inequity are products of design. They can
  be redesigned}.
\newblock
\newblock


\bibitem[Hoffman(2021)]%
        {Hoffman2021}
\bibfield{author}{\bibinfo{person}{Anna~Lauren Hoffman}.}
  \bibinfo{year}{2021}\natexlab{}.
\newblock \showarticletitle{Terms of Inclusion: Data, discourse, violence}.
\newblock \bibinfo{journal}{\emph{New Media \& Society}} \bibinfo{volume}{23},
  \bibinfo{number}{12} (\bibinfo{year}{2021}), \bibinfo{pages}{3539--3556}.
\newblock
\urldef\tempurl%
\url{https://doi.org/10.1177/1461444820958725}
\showDOI{\tempurl}


\bibitem[Hutchinson and Mitchell(2019)]%
        {Hutchinson2019}
\bibfield{author}{\bibinfo{person}{Ben Hutchinson} {and}
  \bibinfo{person}{Margaret Mitchell}.} \bibinfo{year}{2019}\natexlab{}.
\newblock \showarticletitle{50 Years of Test (Un)Fairness: Lessons for Machine
  Learning}. In \bibinfo{booktitle}{\emph{Proceedings of the Conference on
  Fairness, Accountability, and Transparency}} (Atlanta, GA, USA)
  \emph{(\bibinfo{series}{FAT* '19})}. \bibinfo{publisher}{Association for
  Computing Machinery}, \bibinfo{address}{New York, NY, USA},
  \bibinfo{pages}{49–58}.
\newblock
\showISBNx{9781450361255}
\urldef\tempurl%
\url{https://doi.org/10.1145/3287560.3287600}
\showDOI{\tempurl}


\bibitem[Katell et~al\mbox{.}(2020)]%
        {katell2020}
\bibfield{author}{\bibinfo{person}{Michael Katell}, \bibinfo{person}{Meg
  Young}, \bibinfo{person}{Dharma Dailey}, \bibinfo{person}{Bernease Herman},
  \bibinfo{person}{Vivian Guetler}, \bibinfo{person}{Aaron Tam},
  \bibinfo{person}{Corinne Bintz}, \bibinfo{person}{Daniella Raz}, {and}
  \bibinfo{person}{P.~M. Krafft}.} \bibinfo{year}{2020}\natexlab{}.
\newblock \showarticletitle{Toward Situated Interventions for Algorithmic
  Equity: Lessons from the Field}. In \bibinfo{booktitle}{\emph{Proceedings of
  the 2020 Conference on Fairness, Accountability, and Transparency}}
  (Barcelona, Spain) \emph{(\bibinfo{series}{FAT* '20})}.
  \bibinfo{publisher}{Association for Computing Machinery},
  \bibinfo{address}{New York, NY, USA}, \bibinfo{pages}{45–55}.
\newblock
\showISBNx{9781450369367}
\urldef\tempurl%
\url{https://doi.org/10.1145/3351095.3372874}
\showDOI{\tempurl}


\bibitem[Keel(2018)]%
        {Keel2018}
\bibfield{author}{\bibinfo{person}{Terence Keel}.}
  \bibinfo{year}{2018}\natexlab{}.
\newblock \bibinfo{booktitle}{\emph{Divine Variations: How Christian Thought
  Became Racial Science}}.
\newblock \bibinfo{publisher}{Stanford University Press}.
\newblock


\bibitem[Keller and Longino(1996)]%
        {Keller1996}
\bibfield{author}{\bibinfo{person}{Evelyn~Fox Keller} {and}
  \bibinfo{person}{Helen Longino}.} \bibinfo{year}{1996}\natexlab{}.
\newblock \bibinfo{booktitle}{\emph{Feminism and Science}}.
\newblock \bibinfo{publisher}{Oxford University Press}.
\newblock


\bibitem[Klumbyt{\.e} et~al\mbox{.}(2022)]%
        {Klumbyte2022}
\bibfield{author}{\bibinfo{person}{Goda Klumbyt{\.e}}, \bibinfo{person}{Claude
  Draude}, {and} \bibinfo{person}{Alex~S Taylor}.}
  \bibinfo{year}{2022}\natexlab{}.
\newblock \showarticletitle{Critical Tools for Machine Learning: Working with
  Intersectional Critical Concepts in Machine Learning Systems Design}. In
  \bibinfo{booktitle}{\emph{2022 ACM Conference on Fairness, Accountability,
  and Transparency}}. \bibinfo{pages}{1528--1541}.
\newblock


\bibitem[Knorr~Cetina(1999)]%
        {Knorr1999}
\bibfield{author}{\bibinfo{person}{Karen Knorr~Cetina}.}
  \bibinfo{year}{1999}\natexlab{}.
\newblock \bibinfo{booktitle}{\emph{Epistemic Cultures: How the Sciences Make
  Knowledge}}.
\newblock \bibinfo{publisher}{Harvard University Press}.
\newblock


\bibitem[Lab(2020)]%
        {Precarity2020}
\bibfield{author}{\bibinfo{person}{Precarity Lab}.}
  \bibinfo{year}{2020}\natexlab{}.
\newblock \bibinfo{booktitle}{\emph{Technoprecarious}}.
\newblock \bibinfo{publisher}{Goldsmiths Press}.
\newblock


\bibitem[Latour and Woolgar(1986)]%
        {Latour1986}
\bibfield{author}{\bibinfo{person}{Bruno Latour} {and} \bibinfo{person}{Steve
  Woolgar}.} \bibinfo{year}{1986}\natexlab{}.
\newblock \bibinfo{booktitle}{\emph{Laboratory Life: The Construction of
  Scientific Facts}}.
\newblock \bibinfo{publisher}{Princeton University Press}.
\newblock


\bibitem[Laufer et~al\mbox{.}(2022)]%
        {Laufer2022}
\bibfield{author}{\bibinfo{person}{Benjamin Laufer}, \bibinfo{person}{Sameer
  Jain}, \bibinfo{person}{A.~Feder Cooper}, \bibinfo{person}{Jon Kleinberg},
  {and} \bibinfo{person}{Hoda Heidari}.} \bibinfo{year}{2022}\natexlab{}.
\newblock \showarticletitle{Four Years of FAccT: A Reflexive, Mixed-Methods
  Analysis of Research Contributions, Shortcomings, and Future Prospects}. In
  \bibinfo{booktitle}{\emph{2022 ACM Conference on Fairness, Accountability,
  and Transparency}} (Seoul, Republic of Korea) \emph{(\bibinfo{series}{FAccT
  '22})}. \bibinfo{publisher}{Association for Computing Machinery},
  \bibinfo{address}{New York, NY, USA}, \bibinfo{pages}{401–426}.
\newblock
\showISBNx{9781450393522}
\urldef\tempurl%
\url{https://doi.org/10.1145/3531146.3533107}
\showDOI{\tempurl}


\bibitem[Leidner and Kayworth(2006)]%
        {Leidner2006}
\bibfield{author}{\bibinfo{person}{Dorothy~E Leidner} {and}
  \bibinfo{person}{Timothy Kayworth}.} \bibinfo{year}{2006}\natexlab{}.
\newblock \showarticletitle{A review of culture in information systems
  research: Toward a theory of information technology culture conflict}.
\newblock \bibinfo{journal}{\emph{MIS quarterly}} (\bibinfo{year}{2006}),
  \bibinfo{pages}{357--399}.
\newblock


\bibitem[Liao and Sundar(2022)]%
        {Liao2022}
\bibfield{author}{\bibinfo{person}{Q~Vera Liao} {and} \bibinfo{person}{S~Shyam
  Sundar}.} \bibinfo{year}{2022}\natexlab{}.
\newblock \showarticletitle{Designing for Responsible Trust in AI Systems: A
  Communication Perspective}.
\newblock \bibinfo{journal}{\emph{arXiv preprint arXiv:2204.13828}}
  (\bibinfo{year}{2022}).
\newblock


\bibitem[Luccioni et~al\mbox{.}(2022)]%
        {Luccioni2022}
\bibfield{author}{\bibinfo{person}{Alexandra~Sasha Luccioni},
  \bibinfo{person}{Frances Corry}, \bibinfo{person}{Hamsini Sridharan},
  \bibinfo{person}{Mike Ananny}, \bibinfo{person}{Jason Schultz}, {and}
  \bibinfo{person}{Kate Crawford}.} \bibinfo{year}{2022}\natexlab{}.
\newblock \showarticletitle{A Framework for Deprecating Datasets: Standardizing
  Documentation, Identification, and Communication}. In
  \bibinfo{booktitle}{\emph{2022 ACM Conference on Fairness, Accountability,
  and Transparency}}. \bibinfo{pages}{199--212}.
\newblock


\bibitem[Ludwig and Macnaghten(2020)]%
        {Ludwig2020}
\bibfield{author}{\bibinfo{person}{David Ludwig} {and} \bibinfo{person}{Phil
  Macnaghten}.} \bibinfo{year}{2020}\natexlab{}.
\newblock \showarticletitle{Traditional ecological knowledge in innovation
  governance: a framework for responsible and just innovation}.
\newblock \bibinfo{journal}{\emph{Journal of Responsible Innovation}}
  \bibinfo{volume}{7}, \bibinfo{number}{1} (\bibinfo{year}{2020}),
  \bibinfo{pages}{26--44}.
\newblock
\urldef\tempurl%
\url{https://doi.org/10.1080/23299460.2019.1676686}
\showDOI{\tempurl}
\showeprint{https://doi.org/10.1080/23299460.2019.1676686}


\bibitem[Lundberg and Lee(2017)]%
        {Lundberg2017}
\bibfield{author}{\bibinfo{person}{Scott~M. Lundberg} {and}
  \bibinfo{person}{Su-In Lee}.} \bibinfo{year}{2017}\natexlab{}.
\newblock \showarticletitle{A Unified Approach to Interpreting Model
  Predictions}. In \bibinfo{booktitle}{\emph{Proceedings of the 31st
  International Conference on Neural Information Processing Systems}} (Long
  Beach, California, USA) \emph{(\bibinfo{series}{NIPS'17})}.
  \bibinfo{publisher}{Curran Associates Inc.}, \bibinfo{address}{Red Hook, NY,
  USA}, \bibinfo{pages}{4768–4777}.
\newblock
\showISBNx{9781510860964}


\bibitem[Magnet(2011)]%
        {Magnet2011}
\bibfield{author}{\bibinfo{person}{Shoshana~Amielle Magnet}.}
  \bibinfo{year}{2011}\natexlab{}.
\newblock \bibinfo{booktitle}{\emph{When Biometrics Fail: Gender, Race, and the
  Technology of Identity}}.
\newblock \bibinfo{publisher}{Duke University Press}.
\newblock


\bibitem[Maslow(1966)]%
        {Maslow1966}
\bibfield{author}{\bibinfo{person}{Abraham~Harold Maslow}.}
  \bibinfo{year}{1966}\natexlab{}.
\newblock \showarticletitle{The psychology of science a reconnaissance}.
\newblock  (\bibinfo{year}{1966}).
\newblock


\bibitem[Minh-Tam Dao-Kroeker et~al\mbox{.}(2021)]%
        {Minh2021}
\bibfield{author}{\bibinfo{person}{Zoe Minh-Tam Dao-Kroeker},
  \bibinfo{person}{Alexandra Kitson}, \bibinfo{person}{Alissa N.~Antle},
  \bibinfo{person}{Yumiko Murai}, {and} \bibinfo{person}{Azadeh Adibi}.}
  \bibinfo{year}{2021}\natexlab{}.
\newblock \showarticletitle{Designing Biotech ethics cards: Promoting critical
  making during an online workshop with youth}. In
  \bibinfo{booktitle}{\emph{Interaction Design and Children}}.
  \bibinfo{pages}{450--455}.
\newblock


\bibitem[Minna~Stern(2005)]%
        {Stern2005}
\bibfield{author}{\bibinfo{person}{Alexandra Minna~Stern}.}
  \bibinfo{year}{2005}\natexlab{}.
\newblock \bibinfo{booktitle}{\emph{Eugenic Nation: Faults and Frontiers of
  Better Breeding in Modern America}}.
\newblock \bibinfo{publisher}{University of California Press}.
\newblock


\bibitem[Mitchell et~al\mbox{.}(2019)]%
        {Mitchell2019}
\bibfield{author}{\bibinfo{person}{Margaret Mitchell}, \bibinfo{person}{Simone
  Wu}, \bibinfo{person}{Andrew Zaldivar}, \bibinfo{person}{Parker Barnes},
  \bibinfo{person}{Lucy Vasserman}, \bibinfo{person}{Ben Hutchinson},
  \bibinfo{person}{Elena Spitzer}, \bibinfo{person}{Inioluwa~Deborah Raji},
  {and} \bibinfo{person}{Timnit Gebru}.} \bibinfo{year}{2019}\natexlab{}.
\newblock \showarticletitle{Model Cards for Model Reporting}. In
  \bibinfo{booktitle}{\emph{Proceedings of the Conference on Fairness,
  Accountability, and Transparency}} (Atlanta, GA, USA)
  \emph{(\bibinfo{series}{FAT* '19})}. \bibinfo{publisher}{Association for
  Computing Machinery}, \bibinfo{address}{New York, NY, USA},
  \bibinfo{pages}{220–229}.
\newblock
\showISBNx{9781450361255}
\urldef\tempurl%
\url{https://doi.org/10.1145/3287560.3287596}
\showDOI{\tempurl}


\bibitem[Moore(2020)]%
        {Moore2020}
\bibfield{author}{\bibinfo{person}{Jared Moore}.}
  \bibinfo{year}{2020}\natexlab{}.
\newblock \showarticletitle{Towards a More Representative Politics in the
  Ethics of Computer Science}. In \bibinfo{booktitle}{\emph{Proceedings of the
  2020 Conference on Fairness, Accountability, and Transparency}} (Barcelona,
  Spain) \emph{(\bibinfo{series}{FAT* '20})}. \bibinfo{publisher}{Association
  for Computing Machinery}, \bibinfo{address}{New York, NY, USA},
  \bibinfo{pages}{414–424}.
\newblock
\showISBNx{9781450369367}
\urldef\tempurl%
\url{https://doi.org/10.1145/3351095.3372854}
\showDOI{\tempurl}


\bibitem[Morozov(2013)]%
        {Morozov2013}
\bibfield{author}{\bibinfo{person}{Evgeny Morozov}.}
  \bibinfo{year}{2013}\natexlab{}.
\newblock \bibinfo{booktitle}{\emph{To save everything, click here: The folly
  of technological solutionism}}.
\newblock \bibinfo{publisher}{Public Affairs}.
\newblock


\bibitem[Moss and Metcalf(2020)]%
        {Moss2020}
\bibfield{author}{\bibinfo{person}{Emanuel Moss} {and} \bibinfo{person}{Jacob
  Metcalf}.} \bibinfo{year}{2020}\natexlab{}.
\newblock \bibinfo{title}{Ethics Owners: A New Model of Organizational
  Responsibility in Data-Driven Technology Companies}.
\newblock
\newblock
\urldef\tempurl%
\url{https://datasociety.net/pubs/Ethics-Owners.pdf}
\showURL{%
\tempurl}


\bibitem[Nakamura(2007)]%
        {Nakamura2007}
\bibfield{author}{\bibinfo{person}{Lisa Nakamura}.}
  \bibinfo{year}{2007}\natexlab{}.
\newblock \bibinfo{booktitle}{\emph{Digitizing Race: Visual Cultures of the
  Internet}}.
\newblock \bibinfo{publisher}{Minnesota University Press}.
\newblock


\bibitem[Noble(2018)]%
        {Noble2018}
\bibfield{author}{\bibinfo{person}{Safiya~Umoja Noble}.}
  \bibinfo{year}{2018}\natexlab{}.
\newblock \bibinfo{booktitle}{\emph{Algorithms of Oppression: How Search
  Engines Reinforce Racism}}.
\newblock \bibinfo{publisher}{NYU Press}.
\newblock


\bibitem[Passi and Barocas(2019)]%
        {Passi2019}
\bibfield{author}{\bibinfo{person}{Samir Passi} {and} \bibinfo{person}{Solon
  Barocas}.} \bibinfo{year}{2019}\natexlab{}.
\newblock \showarticletitle{Problem Formulation and Fairness}. In
  \bibinfo{booktitle}{\emph{Proceedings of the Conference on Fairness,
  Accountability, and Transparency}} (Atlanta, GA, USA)
  \emph{(\bibinfo{series}{FAT* '19})}. \bibinfo{publisher}{Association for
  Computing Machinery}, \bibinfo{address}{New York, NY, USA},
  \bibinfo{pages}{39–48}.
\newblock
\showISBNx{9781450361255}
\urldef\tempurl%
\url{https://doi.org/10.1145/3287560.3287567}
\showDOI{\tempurl}


\bibitem[Passi and Jackson(2018)]%
        {Passi2018}
\bibfield{author}{\bibinfo{person}{Samir Passi} {and}
  \bibinfo{person}{Steven~J. Jackson}.} \bibinfo{year}{2018}\natexlab{}.
\newblock \showarticletitle{Trust in Data Science: Collaboration, Translation,
  and Accountability in Corporate Data Science Projects}.
\newblock \bibinfo{journal}{\emph{Proc. ACM Hum.-Comput. Interact.}}
  \bibinfo{volume}{2}, \bibinfo{number}{CSCW}, Article \bibinfo{articleno}{136}
  (\bibinfo{date}{nov} \bibinfo{year}{2018}), \bibinfo{numpages}{28}~pages.
\newblock
\urldef\tempurl%
\url{https://doi.org/10.1145/3274405}
\showDOI{\tempurl}


\bibitem[Porter(2020)]%
        {Porter2020}
\bibfield{author}{\bibinfo{person}{Theodore~M. Porter}.}
  \bibinfo{year}{2020}\natexlab{}.
\newblock \bibinfo{booktitle}{\emph{Trust in Numbers: The Pursuit of
  Objectivity in Science and Public Life}}.
\newblock \bibinfo{publisher}{Princeton University Press}.
\newblock


\bibitem[Redfield(2000)]%
        {Redfield2000}
\bibfield{author}{\bibinfo{person}{Peter Redfield}.}
  \bibinfo{year}{2000}\natexlab{}.
\newblock \bibinfo{booktitle}{\emph{Space in the Tropics: From Convicts to
  Rockets in French Guiana}}.
\newblock \bibinfo{publisher}{University of California Press}.
\newblock


\bibitem[Ribeiro et~al\mbox{.}(2016)]%
        {Ribeiro2016}
\bibfield{author}{\bibinfo{person}{Marco~Tulio Ribeiro},
  \bibinfo{person}{Sameer Singh}, {and} \bibinfo{person}{Carlos Guestrin}.}
  \bibinfo{year}{2016}\natexlab{}.
\newblock \showarticletitle{"Why Should I Trust You?": Explaining the
  Predictions of Any Classifier}. In \bibinfo{booktitle}{\emph{Proceedings of
  the 22nd ACM SIGKDD International Conference on Knowledge Discovery and Data
  Mining}} (San Francisco, California, USA) \emph{(\bibinfo{series}{KDD '16})}.
  \bibinfo{publisher}{Association for Computing Machinery},
  \bibinfo{address}{New York, NY, USA}, \bibinfo{pages}{1135–1144}.
\newblock
\showISBNx{9781450342322}
\urldef\tempurl%
\url{https://doi.org/10.1145/2939672.2939778}
\showDOI{\tempurl}


\bibitem[Saja et~al\mbox{.}(2018)]%
        {Saja2018}
\bibfield{author}{\bibinfo{person}{A.M.~Aslam Saja}, \bibinfo{person}{Melissa
  Teo}, \bibinfo{person}{Ashantha Goonetilleke}, {and}
  \bibinfo{person}{Abdul~M. Ziyath}.} \bibinfo{year}{2018}\natexlab{}.
\newblock \showarticletitle{An inclusive and adaptive framework for measuring
  social resilience to disasters}.
\newblock \bibinfo{journal}{\emph{International Journal of Disaster Risk
  Reduction}}  \bibinfo{volume}{28} (\bibinfo{year}{2018}),
  \bibinfo{pages}{862--873}.
\newblock
\showISSN{2212-4209}
\urldef\tempurl%
\url{https://doi.org/10.1016/j.ijdrr.2018.02.004}
\showDOI{\tempurl}


\bibitem[Schiff et~al\mbox{.}(2020)]%
        {Schiff2020}
\bibfield{author}{\bibinfo{person}{Daniel Schiff}, \bibinfo{person}{Bogdana
  Rakova}, \bibinfo{person}{Aladdin Ayesh}, \bibinfo{person}{Anat Fanti}, {and}
  \bibinfo{person}{Michael Lennon}.} \bibinfo{year}{2020}\natexlab{}.
\newblock \bibinfo{title}{Principles to Practices for Responsible AI: Closing
  the Gap}.
\newblock
\newblock
\urldef\tempurl%
\url{https://doi.org/10.48550/ARXIV.2006.04707}
\showDOI{\tempurl}


\bibitem[Schwartz~Cowan(1985)]%
        {cowan1985}
\bibfield{author}{\bibinfo{person}{Ruth Schwartz~Cowan}.}
  \bibinfo{year}{1985}\natexlab{}.
\newblock \showarticletitle{How the refrigerator got its hum}.
\newblock In \bibinfo{booktitle}{\emph{The social shaping of technology}},
  \bibfield{editor}{\bibinfo{person}{Donald MacKenzie} {and}
  \bibinfo{person}{Judy Wajcman}} (Eds.). \bibinfo{publisher}{McGraw Hill
  Education}, \bibinfo{pages}{202--218}.
\newblock


\bibitem[Scott et~al\mbox{.}(2022)]%
        {Scott2022}
\bibfield{author}{\bibinfo{person}{Kristen~M. Scott},
  \bibinfo{person}{Sonja~Mei Wang}, \bibinfo{person}{Milagros Miceli},
  \bibinfo{person}{Pieter Delobelle}, \bibinfo{person}{Karolina
  Sztandar-Sztanderska}, {and} \bibinfo{person}{Bettina Berendt}.}
  \bibinfo{year}{2022}\natexlab{}.
\newblock \showarticletitle{Algorithmic Tools in Public Employment Services:
  Towards a Jobseeker-Centric Perspective}. In \bibinfo{booktitle}{\emph{2022
  ACM Conference on Fairness, Accountability, and Transparency}} (Seoul,
  Republic of Korea) \emph{(\bibinfo{series}{FAccT '22})}.
  \bibinfo{publisher}{Association for Computing Machinery},
  \bibinfo{address}{New York, NY, USA}, \bibinfo{pages}{2138–2148}.
\newblock
\showISBNx{9781450393522}
\urldef\tempurl%
\url{https://doi.org/10.1145/3531146.3534631}
\showDOI{\tempurl}


\bibitem[Selbst et~al\mbox{.}(2019)]%
        {Selbst2019}
\bibfield{author}{\bibinfo{person}{Andrew~D. Selbst}, \bibinfo{person}{Danah
  Boyd}, \bibinfo{person}{Sorelle~A. Friedler}, \bibinfo{person}{Suresh
  Venkatasubramanian}, {and} \bibinfo{person}{Janet Vertesi}.}
  \bibinfo{year}{2019}\natexlab{}.
\newblock \showarticletitle{Fairness and Abstraction in Sociotechnical
  Systems}. In \bibinfo{booktitle}{\emph{Proceedings of the Conference on
  Fairness, Accountability, and Transparency}} (Atlanta, GA, USA)
  \emph{(\bibinfo{series}{FAT* '19})}. \bibinfo{publisher}{Association for
  Computing Machinery}, \bibinfo{address}{New York, NY, USA},
  \bibinfo{pages}{59–68}.
\newblock
\showISBNx{9781450361255}
\urldef\tempurl%
\url{https://doi.org/10.1145/3287560.3287598}
\showDOI{\tempurl}


\bibitem[Shapin and Schaffer(2017)]%
        {Shapin2017}
\bibfield{author}{\bibinfo{person}{Seven Shapin} {and} \bibinfo{person}{Simon
  Schaffer}.} \bibinfo{year}{2017}\natexlab{}.
\newblock \bibinfo{booktitle}{\emph{Leviathan and the Air-Pump: Hobbes, Boyle,
  and the Experimental Life}}.
\newblock \bibinfo{publisher}{Princeton University Press}.
\newblock


\bibitem[Shepherd and Ahmed(2000)]%
        {Shepherd2000}
\bibfield{author}{\bibinfo{person}{Charles Shepherd} {and}
  \bibinfo{person}{Pervaiz~K Ahmed}.} \bibinfo{year}{2000}\natexlab{}.
\newblock \showarticletitle{From product innovation to solutions innovation: a
  new paradigm for competitive advantage}.
\newblock \bibinfo{journal}{\emph{European Journal of Innovation Management}}
  (\bibinfo{year}{2000}).
\newblock


\bibitem[Shneiderman(2021)]%
        {Shneiderman2021}
\bibfield{author}{\bibinfo{person}{Ben Shneiderman}.}
  \bibinfo{year}{2021}\natexlab{}.
\newblock \showarticletitle{Responsible AI: Bridging from Ethics to Practice}.
\newblock \bibinfo{journal}{\emph{Commun. ACM}} \bibinfo{volume}{64},
  \bibinfo{number}{8} (\bibinfo{date}{jul} \bibinfo{year}{2021}),
  \bibinfo{pages}{32–35}.
\newblock
\showISSN{0001-0782}
\urldef\tempurl%
\url{https://doi.org/10.1145/3445973}
\showDOI{\tempurl}


\bibitem[Smyth and Dimond(2014)]%
        {Smyth2014}
\bibfield{author}{\bibinfo{person}{Thomas Smyth} {and} \bibinfo{person}{Jill
  Dimond}.} \bibinfo{year}{2014}\natexlab{}.
\newblock \showarticletitle{Anti-oppressive design}.
\newblock \bibinfo{journal}{\emph{Interactions}} \bibinfo{volume}{21},
  \bibinfo{number}{6} (\bibinfo{year}{2014}), \bibinfo{pages}{68--71}.
\newblock


\bibitem[Stamper(1993)]%
        {Stamper1993}
\bibfield{author}{\bibinfo{person}{Ronald Stamper}.}
  \bibinfo{year}{1993}\natexlab{}.
\newblock \showarticletitle{A semiotic theory of information and information
  systems}. In \bibinfo{booktitle}{\emph{Invited papers for the ICL/University
  of Newcastle Seminar on Information}}.
\newblock


\bibitem[Stark(2009)]%
        {stark2009}
\bibfield{author}{\bibinfo{person}{David Stark}.}
  \bibinfo{year}{2009}\natexlab{}.
\newblock \bibinfo{booktitle}{\emph{The Sense of Dissonance: Accounts of Worth
  in Economic Life}}.
\newblock \bibinfo{publisher}{Princeton University Press}.
\newblock


\bibitem[Stilgoe et~al\mbox{.}(2013)]%
        {Stilgoe2013}
\bibfield{author}{\bibinfo{person}{Jack Stilgoe}, \bibinfo{person}{Richard
  Owen}, {and} \bibinfo{person}{Phil Macnaghten}.}
  \bibinfo{year}{2013}\natexlab{}.
\newblock \showarticletitle{Developing a framework for responsible innovation}.
\newblock \bibinfo{journal}{\emph{Research Policy}} \bibinfo{volume}{42},
  \bibinfo{number}{9} (\bibinfo{year}{2013}), \bibinfo{pages}{1568--1580}.
\newblock
\showISSN{0048-7333}
\urldef\tempurl%
\url{https://doi.org/10.1016/j.respol.2013.05.008}
\showDOI{\tempurl}


\bibitem[TallBear(2013)]%
        {TallBear2013}
\bibfield{author}{\bibinfo{person}{Kim TallBear}.}
  \bibinfo{year}{2013}\natexlab{}.
\newblock \bibinfo{booktitle}{\emph{Native American DNA: Tribal Belonging and
  the False Promise of Genetic Science}}.
\newblock \bibinfo{publisher}{Minnesota University Press}.
\newblock


\bibitem[Terzis(2020)]%
        {Terzis2020}
\bibfield{author}{\bibinfo{person}{Petros Terzis}.}
  \bibinfo{year}{2020}\natexlab{}.
\newblock \showarticletitle{Onward for the Freedom of Others: Marching beyond
  the AI Ethics}. In \bibinfo{booktitle}{\emph{Proceedings of the 2020
  Conference on Fairness, Accountability, and Transparency}} (Barcelona, Spain)
  \emph{(\bibinfo{series}{FAT* '20})}. \bibinfo{publisher}{Association for
  Computing Machinery}, \bibinfo{address}{New York, NY, USA},
  \bibinfo{pages}{220–229}.
\newblock
\showISBNx{9781450369367}
\urldef\tempurl%
\url{https://doi.org/10.1145/3351095.3373152}
\showDOI{\tempurl}


\bibitem[Tomasini~Giannini and Mulder(2022)]%
        {Tomasini_Giannini2022-dd}
\bibfield{author}{\bibinfo{person}{Fabiana Tomasini~Giannini} {and}
  \bibinfo{person}{Ingrid Mulder}.} \bibinfo{year}{2022}\natexlab{}.
\newblock \showarticletitle{Towards a {Power-Balanced} Participatory Design
  Process}. In \bibinfo{booktitle}{\emph{Proceedings of the Participatory
  Design Conference 2022 - Volume 2}} (Newcastle upon Tyne, United Kingdom)
  \emph{(\bibinfo{series}{PDC '22})}. \bibinfo{publisher}{Association for
  Computing Machinery}, \bibinfo{address}{New York, NY, USA},
  \bibinfo{pages}{111--117}.
\newblock


\bibitem[van Oorschot et~al\mbox{.}(2022)]%
        {Van_Oorschot2022-ss}
\bibfield{author}{\bibinfo{person}{Robin van Oorschot}, \bibinfo{person}{Dirk
  Snelders}, \bibinfo{person}{Maaike Kleinsmann}, {and} \bibinfo{person}{Jacob
  Buur}.} \bibinfo{year}{2022}\natexlab{}.
\newblock \showarticletitle{Participation in design research}.
\newblock \bibinfo{journal}{\emph{Design Studies}}  \bibinfo{volume}{78}
  (\bibinfo{date}{Jan.} \bibinfo{year}{2022}), \bibinfo{pages}{101073}.
\newblock


\bibitem[Varnum and Grossmann(2017)]%
        {Varnum2017}
\bibfield{author}{\bibinfo{person}{Michael E.~W. Varnum} {and}
  \bibinfo{person}{Igor Grossmann}.} \bibinfo{year}{2017}\natexlab{}.
\newblock \showarticletitle{Cultural Change: The How and the Why}.
\newblock \bibinfo{journal}{\emph{Perspectives on Psychological Science}}
  \bibinfo{volume}{12}, \bibinfo{number}{6} (\bibinfo{year}{2017}),
  \bibinfo{pages}{956--972}.
\newblock
\urldef\tempurl%
\url{https://doi.org/10.1177/1745691617699971}
\showDOI{\tempurl}
\showeprint{https://doi.org/10.1177/1745691617699971}
\newblock
\shownote{PMID: 28915361}.


\bibitem[Voegtlin and Scherer(2017)]%
        {Voegtlin2017}
\bibfield{author}{\bibinfo{person}{Christian Voegtlin} {and}
  \bibinfo{person}{Andreas~Georg Scherer}.} \bibinfo{year}{2017}\natexlab{}.
\newblock \showarticletitle{Responsible innovation and the innovation of
  responsibility: Governing sustainable development in a globalized world}.
\newblock \bibinfo{journal}{\emph{Journal of business ethics}}
  \bibinfo{volume}{143}, \bibinfo{number}{2} (\bibinfo{year}{2017}),
  \bibinfo{pages}{227--243}.
\newblock


\bibitem[Voida et~al\mbox{.}(2014)]%
        {voida_2014}
\bibfield{author}{\bibinfo{person}{Amy Voida}, \bibinfo{person}{Lynn
  Dombrowski}, \bibinfo{person}{Gillian~R. Hayes}, {and}
  \bibinfo{person}{Melissa Mazmanian}.} \bibinfo{year}{2014}\natexlab{}.
\newblock \showarticletitle{Shared Values/Conflicting Logics: Working around
  e-Government Systems}. In \bibinfo{booktitle}{\emph{Proceedings of the SIGCHI
  Conference on Human Factors in Computing Systems}} (Toronto, Ontario, Canada)
  \emph{(\bibinfo{series}{CHI '14})}. \bibinfo{publisher}{Association for
  Computing Machinery}, \bibinfo{address}{New York, NY, USA},
  \bibinfo{pages}{3583–3592}.
\newblock
\showISBNx{9781450324731}
\urldef\tempurl%
\url{https://doi.org/10.1145/2556288.2556971}
\showDOI{\tempurl}


\bibitem[Watts(2017)]%
        {Watts}
\bibfield{author}{\bibinfo{person}{D Watts}.} \bibinfo{year}{2017}\natexlab{}.
\newblock \showarticletitle{Should social science be more solution-oriented?}
\newblock \bibinfo{journal}{\emph{Nature Human Behavior}}
  (\bibinfo{year}{2017}).
\newblock
\urldef\tempurl%
\url{https://doi.org/10.1038/s41562-016-0015}
\showDOI{\tempurl}


\bibitem[Wetzel et~al\mbox{.}(2017)]%
        {Wetzel2017}
\bibfield{author}{\bibinfo{person}{Richard Wetzel}, \bibinfo{person}{Tom
  Rodden}, {and} \bibinfo{person}{Steve Benford}.}
  \bibinfo{year}{2017}\natexlab{}.
\newblock \showarticletitle{Developing ideation cards for mixed reality game
  design}.
\newblock \bibinfo{journal}{\emph{Transactions of the Digital Games Research
  Association}} \bibinfo{volume}{3}, \bibinfo{number}{2}
  (\bibinfo{year}{2017}).
\newblock


\bibitem[Whittlestone et~al\mbox{.}(2019)]%
        {Whittlestone2019}
\bibfield{author}{\bibinfo{person}{Jess Whittlestone}, \bibinfo{person}{Rune
  Nyrup}, \bibinfo{person}{Anna Alexandrova}, {and} \bibinfo{person}{Stephen
  Cave}.} \bibinfo{year}{2019}\natexlab{}.
\newblock \showarticletitle{The Role and Limits of Principles in AI Ethics:
  Towards a Focus on Tensions}. In \bibinfo{booktitle}{\emph{Proceedings of the
  2019 AAAI/ACM Conference on AI, Ethics, and Society}} (Honolulu, HI, USA)
  \emph{(\bibinfo{series}{AIES '19})}. \bibinfo{publisher}{Association for
  Computing Machinery}, \bibinfo{address}{New York, NY, USA},
  \bibinfo{pages}{195–200}.
\newblock
\showISBNx{9781450363242}
\urldef\tempurl%
\url{https://doi.org/10.1145/3306618.3314289}
\showDOI{\tempurl}


\bibitem[Winner(1980)]%
        {Winner1980}
\bibfield{author}{\bibinfo{person}{Langdon Winner}.}
  \bibinfo{year}{1980}\natexlab{}.
\newblock \showarticletitle{Do artifacts have politics?}
\newblock \bibinfo{journal}{\emph{Deadalus}} (\bibinfo{year}{1980}),
  \bibinfo{pages}{121--136}.
\newblock


\bibitem[Woodings(2006)]%
        {Woodings2006}
\bibfield{author}{\bibinfo{person}{Bruce Woodings}.}
  \bibinfo{year}{2006}\natexlab{}.
\newblock \showarticletitle{Aligning large multi-cultural teams performance
  with a solutions focused approach}.
\newblock In \bibinfo{booktitle}{\emph{Solution-focused management}}.
  \bibinfo{publisher}{Reiner Hampp Verlag}, \bibinfo{pages}{371--382}.
\newblock


\bibitem[Zuboff(2019)]%
        {Zuboff2019}
\bibfield{author}{\bibinfo{person}{Shoshana Zuboff}.}
  \bibinfo{year}{2019}\natexlab{}.
\newblock \bibinfo{booktitle}{\emph{The Age of Surveillance Capitalism: The
  Fight for a Human Future at the New Frontier of Power}}.
\newblock \bibinfo{publisher}{Public Affairs}.
\newblock


\end{thebibliography}

\appendix

\end{document}